\documentclass[12pt]{article}
\usepackage{epsfig}
\usepackage{axodraw}
\date{\today}
\newcommand{\bmat}{\left(\begin{array}}
\newcommand{\emat}{\end{array}\right)}
\newcommand{\be}{\begin{equation}}
\newcommand{\ee}{\end{equation}}
\newcommand{\ba}{\begin{eqnarray}}
\newcommand{\ea}{\end{eqnarray}}

\def\lsim{\raise0.3ex\hbox{$\;<$\kern-0.75em\raise-1.1ex\hbox{$\sim\;$}}}
\def\gsim{\raise0.3ex\hbox{$\;>$\kern-0.75em\raise-1.1ex\hbox{$\sim\;$}}}

\def\be{\beta}

\def\lsim{\raise0.3ex\hbox{$\;<$\kern-0.75em\raise-1.1ex\hbox{$\sim\;$}}}
\def\gsim{\raise0.3ex\hbox{$\;>$\kern-0.75em\raise-1.1ex\hbox{$\sim\;$}}}

\begin{document}

\renewcommand{\thefootnote}{\fnsymbol{footnote}}

\pagestyle{empty}
\rightline{SUSX-TH/01-025}
\vskip 1cm
\begin{center}
{\bf \large{ Supersymmetry and Electroweak Leptonic Observables\\[10mm]}}
{Oleg Lebedev$^{1}$ and Will Loinaz$^{2}$\\[6mm]}
\small{$^1$Centre for Theoretical Physics, University of Sussex, Brighton BN1
9QJ,~~UK\\[4mm]}
\small{$^2$Department of Physics, Amherst College, Amherst MA 01002,~~ USA
\\[4mm]}
\end{center}
\hrule
\vskip 0.3cm
\begin{minipage}[h]{14.0cm}
\begin{center}
\small{\bf Abstract}\\[3mm]
\end{center}

We study vertex corrections to the leptonic electroweak 
observables in the general MSSM at $\tan\beta \lsim 35$. 
In particular, we address the question of whether 
supersymmetry can be responsible for the observed $2\sigma$ deviation 
from the Standard Model
prediction in the invisible width of the $Z$.
We find that the presence of a light (around 100 GeV) chargino 
and sleptons hinted by the $g_\mu$-2 measurements makes the agreement with 
experiment slightly  better and improves the electroweak fit.

\end{minipage}
\vskip 0.2cm
\hrule
\vskip 1cm

\section{Introduction}

The recent BNL measurements of the muon anomalous magnetic moment
have bolstered interest in supersymmetric models \cite{Brown:2001mg}. 
These measurements appear to deviate from Standard Model (SM) predictions
by 2.6$\sigma$ \cite{Czarnecki:2001pv}.
A conclusive statement can be made only after sufficient statistics have 
been accumulated and 
the status of the SM theoretical uncertainties has been determined unambiguously 
\cite{Yndurain:2001qw}.
However, should this deviation persist and its error shrink, new physics would
be required to explain it.  Among the candidate models for new physics, 
supersymmetric models seem most promising \cite{Czarnecki:2001pv}.

In addition to this possible deviation, there are a number of  
other discrepancies of
similar size between the SM predictions and the experimental values of 
the electroweak observables. In particular, there is
a  more  long-standing $A_b$ ``anomaly'' \cite{Chanowitz:1999jj},\cite{Swartz:2000xv} 
which manifests in
a $2.7\sigma$ deviation of the combined left-right asymmetry in $Z\rightarrow b \bar b$
decays measured at LEP and SLD from the SM prediction. It has been argued that 
such a discrepancy is unlikely to be a result of a statistical deviation 
(see e.g. the work of Chanowitz in Ref.\cite{Chanowitz:1999jj}). 
The possibility of supersymmetric origin of this ``anomaly'' will be pursued
in a subsequent paper.
In addition,  there is a $2\sigma$ deviation in the invisible width of the
Z boson \cite{Swartz:2000xv},
which appears as a deviation of the effective number of neutrinos from three:
\begin{equation}
N_\nu = 2.9835 \pm 0.0083\;.
\end{equation}
Implications of these results for various models of new physics have been considered
in Refs.\cite{Lebedev:2000ze}-\cite{Chang:2001xy}.
In particular, it was found that models with R-parity violating interactions 
\cite{Lebedev:2000ze},
two Higgs doublet models at large $\tan\beta$
 \cite{Lebedev:2000ix},  models with large extra dimensions \cite{Chang:2000yw},
and models with an extra gauge  $U(1)_{B-3L}$ \cite{Chang:2001xy} 
not only fail to mitigate but in fact exacerbate the problem by generating
radiative corrections of the ``wrong'' sign. This observation has 
resulted in stringent constraints on such models.

In this study we analyze the effect of R-conserving
supersymmetry on electroweak leptonic observables and, in particular, 
the invisible width of the Z boson.  Motivated by the supersymmetric explanation
of the BNL $g$-2 ``anomaly'', our study is focused on the question whether or not
the Z invisible width ``anomaly'' can be explained with the same mechanism.
In our analysis we perform a global fit to all relevant electroweak leptonic
observables such as the R-parameters
\begin{eqnarray}
&&R_l=\frac{\Gamma(Z\rightarrow {\rm hadrons})}{\Gamma(Z\rightarrow l^+l^-)}=
\frac{N_c \sum (h^2_{q_L}+h^2_{q_R})}{h^2_{l_L}+h^2_{l_R}}\;,\nonumber\\
&& R_{\nu / e}=\frac{\Gamma(Z\rightarrow \nu \bar \nu)}{\Gamma(Z\rightarrow e^+e^-)}=
\frac{h^2_{\nu_L}}{h^2_{e_L}+h^2_{e_R}}\;,
\label{R}
\end{eqnarray}
the left-right asymmetries
\begin{equation}
A_l= \frac{ h^2_{l_L}-h^2_{l_R}}{h^2_{l_L}+h^2_{l_R}}\;,
\end{equation}
and the forward-backward asymmetries
\begin{equation}
A_{FB}(l)= {3\over 4} A_e A_l\;,
\end{equation}
where $h_{l_{L,R}}$ are the $Z\bar l_{L,R} l_{L,R} $ couplings and $l=e,~\mu,~\tau$. 
The $2\sigma$ deviation in $R_{\nu / e}$ is related to the $\sim2\sigma$ deviation
in $\sigma_{{\rm had}}=12\pi \Gamma_e \Gamma_h / m_Z^2 \Gamma_Z^2$.
We remark that $R_{\nu / e}$ is not measured directly but rather 
calculated from the Z-line-shape observables.
 In principle, $R_{\nu / e}$ could be
affected by SUSY contributions to $\Gamma(Z\rightarrow {\rm hadrons})$; for example,
a light bottom squark may improve agreement with experiment \cite{Carena:2000ka}.
In this work, we concentrate exclusively on the leptonic sector. 

We isolate the effect of the vertex corrections which are 
sensitive to the lepton/chargino-neutralino sector of the MSSM.
The oblique corrections are parameterized in our fit but not used to 
constrain the model due to their significant model-dependence, e.g. they
depend sensitively on the Higgs sector, squark masses, etc. 
In addition, we let $\alpha_s(M_Z)$ float in our fit since its SM value
is extracted from $R_l$.
This strategy has been
used previously and proven useful in placing generic constraints on 
complicated models of new physics 
\cite{Lebedev:2000ze}-\cite{Chang:2001xy}. 
We incorporate the electroweak data reported during summer 2000 conferences
in our numerical analysis.

We present general formulae for the vertex corrections in terms of the
low-energy quantities such as the chargino masses and mixings, 
left and right slepton masses, etc.  We then impose the condition of the 
radiative electroweak symmetry breaking and analyze the GUT scale MSSM  
parameters which improve the electroweak fit.  However, we stress that our 
conclusions are independent
of the assumptions about the high energy structure of the theory and can be
formulated purely in terms of the low energy quantities. 

An analysis which addresses a somewhat similar question but with an emphasis
on the effect of the oblique corrections has recently appeared in Ref.\cite{kaoru}.
We also find a partial overlap with  earlier work \cite{ew}.  Earlier 
calculations of one-loop vertex corrections in the MSSM may be found, e.g.,
 in Ref.\cite{zbb}.

The paper is organized as follows. In Section 2 we present our SUSY framework.
In section 3 we calculate the supersymmetric vertex corrections and study the decoupling
behavior of the SUSY contributions. In section 4 we discuss the fit and our
numerical results, and in sectio 5 we make concluding remarks. 
In the Appendix we list our conventions and relevant Passarino-Veltman functions.

\section{Supersymmetric Framework}

We will study supersymmetric models with the following  superpotential
\begin{equation}
W=-\hat{H}_2\hat{Q}_iY_u^{ij}\hat{U}_j 
+\hat{H}_1\hat{Q}_iY_d^{ij}\hat{D}_j 
+\hat{H}_1\hat{L}_iY_e^{ij}\hat{E}_j-\mu\hat{H}_1\hat{H}_2
\end{equation}
and the high energy scale soft breaking 
potential
\begin{eqnarray}
 V_{SB} &=& \left( m_{0\alpha}^{L}\right)^2 \phi^{L\dagger}_{\alpha} \phi^L_{\alpha} + 
\left( m_{0\alpha}^{R} \right)^2
 \phi^{R*}_{\alpha} \phi^R_{\alpha}  -( B\mu H_1 H_2 +h.c.)
+ \Bigl( A_l Y_{ij}^l\; H_1 \tilde l_{Li} \tilde e_{Rj}^* 
\label{soft}  \\
&+& A_d Y_{ij}^d \; H_1 \tilde q_{Li} \tilde d_{Rj}^* 
- A_u Y_{ij}^u \;  H_2 \tilde q_{Li} \tilde u_{Rj}^* +h.c.\Bigr) 
- {1\over 2} \left( M_3 \lambda^c_3 \lambda^c_3 + M_2 \lambda^a_2 \lambda^a_2
+ M_1 \lambda_1 \lambda_1 \right) \;, \nonumber
\end{eqnarray}
where $\phi_{\alpha}^{L(R)}$ denotes all the scalars of the theory which 
transform under SU(2) as doublets (singlets). We generally allow for 
nonuniversal gaugino and scalar masses. Note that as a result of the SU(2) symmetry
different isospin components of the doublets have the same soft masses at
the high energy scale, whereas there is no similar requirement for the singlets.
At low energies this degeneracy will be broken by the electroweak effects.

In what follows we use $\tan\beta$, $m_{0\alpha}$, $A_{\alpha}$, $M_i$ as input parameters and obtain
low energy quantities via the MSSM renormalization group equations (RGE)
given in Ref.\cite{Bertolini:1991if}.
We also  assume radiative electroweak symmetry breaking, i.e. that the magnitude
of the $\mu$ parameter is given (at tree level) by 
\begin{equation}
\vert \mu \vert^2 = {m^2_{H_1}-m^2_{H_2} \tan^2\beta \over \tan^2\beta -1} - {1\over 2} m_Z^2    \;.
\end{equation}
The phase of  $\mu$ ($\phi_\mu$) is an input parameter and is RG-invariant.

At low energies the charged gauginos and higgsinos mix, leading to the
following mass matrix (we follow the conventions of Ref.\cite{Haber:1985rc};
however, we  correct their sign error in the superpotential)
\begin{eqnarray}
M_{\chi^+}=\left( \matrix{ M_2 & \sqrt{2} M_W \sin\beta \cr
                           \sqrt{2} M_W \cos\beta & \mu}  \right)\;. \nonumber
\end{eqnarray}
This  matrix  is diagonalized by a biunitary transformation
\begin{equation}
U^*~M_{\chi^+}~V^{-1}={\rm diag}(m_{\chi^+_1},m_{\chi^+_2})\;,
\end{equation} 
where $U$ and $V$ are unitary matrices. The mass eigenvalues are defined to be
non-negative  and $m_{\chi^+_1} \geq m_{\chi^+_2}$.

Similarly, for neutralinos we have
\begin{eqnarray}
M_{\chi^0}=\left( \matrix{ M_1 &0 &-M_Z \sin\theta_W \cos\beta & M_Z \sin\theta_W \sin\beta \cr
                           0& M_2 &  M_Z \cos\theta_W \cos\beta &-M_Z \cos\theta_W \sin\beta \cr
                           -M_Z \sin\theta_W \cos\beta &M_Z \cos\theta_W \cos\beta &0&-\mu \cr 
                           M_Z \sin\theta_W \sin\beta &-M_Z \cos\theta_W \sin\beta &-\mu &0} 
 \right)\;.\nonumber
\end{eqnarray}
This symmetric matrix is diagonalized by a unitary matrix $N$,
\begin{equation}
N^*~M_{\chi^0}~N^{-1}={\rm diag}(m_{\chi^0_1},m_{\chi^0_2},m_{\chi^0_3},m_{\chi^0_4})\;,
\end{equation} 
where again the eigenvalues are defined to be non-negative and 
$m_{\chi^0_1} \geq m_{\chi^0_2}\geq ..$. The chargino and neutralino spinors can be split
into the left and right components in the usual way:
\begin{eqnarray}
&& \chi^+_i= \left( \matrix{\chi^+_i \cr
                          {\bar \chi^-_i}}\right) \;\;,\;\;
\chi^0_i= \left( \matrix{\chi^0_i \cr
                          {\bar \chi^0_i }}\right) \;.
\end{eqnarray}

Concerning the slepton spectrum, the ``left'' and ``right'' charged sleptons also mix at low energies.
However, their mixing is proportional to the lepton masses and is negligible unless $\tan\beta$
is very large.  Neglecting lepton masses, the low energy mass eigenstates are:
\begin{eqnarray}
&& m^2_{\tilde e_L}\simeq m^2_{\tilde l}+M_Z^2 \left(-{1\over 2} +\sin^2\theta_W  \right) \cos 2\beta \;,
\nonumber\\
&& m^2_{\tilde e_R}\simeq m^2_{\tilde e}-M_Z^2 \sin^2\theta_W  \cos 2\beta \;, \nonumber\\
&& m^2_{\tilde \nu}\simeq m^2_{\tilde l}+{1\over 2} M_Z^2   \cos 2\beta \;,
\end{eqnarray}
where $m^2_{\tilde l}$ and $m^2_{\tilde e}$ are the mass parameters appearing in the low
energy analog of Eq.(\ref{soft}).

\section{SUSY Vertex Corrections}

In this paper we will concentrate on  $\tan\beta \lsim 35$. It is quite difficult to achieve
radiative electroweak symmetry breaking for greater $\tan\beta$, so such an assumption can
be justified. 
At $\tan\beta \lsim 35$ the gauge couplings dominate the lepton Yukawa couplings, so only the
gaugino parts of the charginos and neutralinos couple to leptons with an appreciable strength.
In addition, one can neglect the left-right slepton mixing in this regime (as will be clear below,
each relevant diagram would require two left-right  mass insertions, so the effect of this mixing
is further suppressed).

We perform our calculations using the two-component spinor technique
(see the Appendix for the notation and conventions).
The result is expressed as a correction $\delta h_{f_{L,R}}$ to the tree level coupling
$h_{f_{L,R}}$ defined by
\begin{equation}
{\cal L}=-{g \over \cos\theta_W} Z_\mu \left[ h_{f_L}~f^\dagger_L \bar\sigma^\mu f_L
+ h_{f_R}~f^\dagger_R \sigma^\mu f_R \right]\;,
\end{equation}
with
\begin{eqnarray}
&& h_{f_L}=I_3-Q\sin^2\theta_W \;,\nonumber\\
&& h_{f_R}=-Q\sin^2\theta_W \;.
\end{eqnarray}

Neglecting the lepton Yukawa couplings, we have the following  SUSY
interactions\footnote{This corrects  an error in Ref.\cite{Haber:1985rc} in the expression for the neutralino
coupling  to right-handed leptons (C77), i.e. $N^*_{j2}$ should be $N^*_{j1}$.}\cite{Haber:1985rc} 
\begin{eqnarray}
 {\cal L}_{l\tilde l \chi^+ }&=&-g\left[ 
(U^*_{11} {\overline {\chi^+_1}} +U^*_{21} {\overline {\chi^+_2}})P_L\nu~\tilde e^*_L +
(V^*_{11} {\overline {\chi^{+c}_1}} +V^*_{21} {\overline {\chi^{+c}_2}})P_L e~\tilde \nu^*
\right]~+~h.c.\;, \nonumber\\
{\cal L}_{l \tilde l\chi^0 }&=& -\sqrt{2} g \sum_j \bar l P_R \chi^0_j ~\tilde l_L
\left[I_3 N_{j2} -\tan\theta_W (I_3-Q) N_{j1} \right] \nonumber\\
 &+& \sqrt{2} g \tan\theta_W \sum_j \bar l P_L \chi^0_j ~ \tilde l_R ~QN^*_{j1}~ +~h.c.\;,\nonumber\\
 {\cal L}_{Z \chi^+ \chi^-  }&=& {g\over \cos\theta_W} Z_\mu \sum_{ij}
   {\overline {\chi^+_i}} \gamma^\mu \left( O_{ij}^{'L} P_L + O_{ij}^{'R} P_R   \right) {\chi^+_j}\;,\nonumber\\
{\cal L}_{Z \chi^0 \chi^0  }&=& {g\over 2\cos\theta_W} Z_\mu \sum_{ij}
   {\overline {\chi^0_i}} \gamma^\mu \left( O_{ij}^{''L} P_L + O_{ij}^{''R} P_R   \right) {\chi^0_j}\;,\nonumber\\
{\cal L}_{Z\tilde l \tilde l  } &=& -{ig\over \cos\theta_W} Z_\mu (I_3-Q\sin^2\theta_W)~
\tilde l^* \stackrel{\leftrightarrow}{\partial^\mu} \tilde l \;.
\end{eqnarray}
Here $I_3$ and $Q$ are the lepton isospin and charge, respectively, and the superscript $c$
stands for a charge conjugated spinor. The vertex structures $O_{ij}$ 
are given by
\begin{eqnarray}
&& O_{ij}^{'L}=-V_{i1}V_{j1}^* -{1\over 2} V_{i2} V_{j2}^* + \delta_{ij} \sin^2\theta_W \;,\nonumber\\
&& O_{ij}^{'R}=-U_{i1}^*U_{j1} -{1\over 2} U_{i2}^* U_{j2} + \delta_{ij} \sin^2\theta_W \;,\nonumber\\
&& O_{ij}^{''L}= -{1\over 2} N_{i3} N_{j3}^* +{1\over 2} N_{i4} N_{j4}^* \;,\nonumber\\
&& O_{ij}^{''R}= -O_{ij}^{''L*}= -O_{ji}^{''L}\;.
\label{O}
\end{eqnarray}
These interactions are to be expressed in terms of the two-component spinors. 
The implementation is trivial
for all interactions except for ${\cal L}_{l\tilde l \chi^+ }$, 
which becomes
\begin{eqnarray}
{\cal L}_{l\tilde l \chi^+ }=-g\left[ 
\nu_L^\dagger \left(U_{11}\chi^+_{1R} +U_{21}\chi^+_{2R}\right)\tilde e_L +
e_L^\dagger~i\sigma_2 \left(V_{11}(\chi^+_{1L})^* +V_{21}(\chi^+_{2L})^*\right) \tilde \nu_L
\right]~+~h.c.\;
\end{eqnarray}
We remark  that $\chi^+$ denotes a Dirac spinor with a positive charge 
(not to be confused with a hermitian conjugated spinor). 

The Z-$\chi^0$-$\chi^0$ coupling can be simplified by 
taking advantage of the Majorana nature of
the neutralino. For Majorana spinors $\psi_1$ and $\psi_2$ we have
\begin{equation}
\overline {\psi_1} \gamma_\mu P_L \psi_2 =- \overline {\psi_2} \gamma_\mu P_R \psi_1 \;.
\end{equation} 
Using this identity as well as $O_{ij}^{''R}= -O_{ji}^{''L}$, we obtain
\begin{eqnarray} 
{\cal L}_{Z \chi^0 \chi^0  }&=& {g\over \cos\theta_W} Z_\mu \sum_{ij}
   {\bar \chi^0_{iR}}~ \sigma^\mu  O_{ij}^{''R} ~{\chi^0_{jR}}\; \nonumber\\
                        &=& {g\over \cos\theta_W} Z_\mu \sum_{ij}
   {\bar \chi^0_{iL}}~ \bar \sigma^\mu  O_{ij}^{''L}~ {\chi^0_{jL}}\;.
\end{eqnarray}

\subsection{Chargino Contributions}

In this subsection we list expressions for Feynman diagrams containing charginos in the loop.
Since the higgsino coupling to leptons can be neglected at $\tan\beta \lsim 35$, the
charginos induce corrections to the left-handed couplings only.  
Below we present our results in terms of the  corrections to the tree level Z-$f_L$-$f_L$ couplings 
$h_{f_L}$.
\begin{eqnarray}
&& \delta h_{e_L}':\nonumber\\
&&(1a): \; g^2 \sum_{ij} O_{ij}^{'L} V_{i1}^* V_{j1}\left[(2-d)\hat C_{24} 
+M_Z^2\hat C_{23}\right](M_Z^2;m_{\tilde \nu}, m_{\chi^+_i},m_{\chi^+_j}) \;, \nonumber\\
&&(1b): \; g^2 \sum_{ij} O_{ij}^{'R} V_{i1}^* V_{j1}~ m_{\chi^+_i} m_{\chi^+_j} 
\hat C_0(M_Z^2;m_{\tilde \nu}, m_{\chi^+_i},m_{\chi^+_j}) \;, \nonumber\\
&&(1c): \;- g^2 \sum_{k} \vert V_{k1} \vert^2~  
\hat C_{24}(M_Z^2;m_{\chi^+_k},m_{\tilde \nu}, m_{\tilde \nu}) \;, \nonumber\\
&&(1d): \;- g^2 \left(-{1\over2}+\sin^2\theta_W\right) \sum_{k} \vert V_{k1} \vert^2~  
B_1(0;m_{\chi^+_k},m_{\tilde \nu})\;.
\label{char.corr.}
\end{eqnarray}
Definitions of the $B$ and $\hat C$ functions can be found in Appendix B. 
We note that  in addition to Fig. 1d there is another wave function renormalization diagram
with the loop on the outgoing electron leg. The corresponding correction is the same
as for the diagram in Fig. 1d, so we do not list it separately. The contribution of the
wave function renormalization diagrams to the total correction comes with a factor of 
1/2, so in effect the total correction is simply given by a sum of individual contributions
in Eq.(\ref{char.corr.}).
The analogous contribution to the (left-handed) neutrino final state is
\begin{eqnarray}
&& \delta h_{\nu}':\nonumber\\
&&(2a): \;- g^2 \sum_{ij} O_{ij}^{'R} U_{j1}^* U_{i1}\left[(2-d)\hat C_{24} 
+M_Z^2\hat C_{23}\right](M_Z^2;m_{\tilde e_L}, m_{\chi^+_j},m_{\chi^+_i}) \;, \nonumber\\
&&(2b): \; -g^2 \sum_{ij} O_{ij}^{'L} U_{j1}^* U_{i1}~ m_{\chi^+_i} m_{\chi^+_j} 
\hat C_0(M_Z^2;m_{\tilde e_L}, m_{\chi^+_j},m_{\chi^+_i}) \;, \nonumber\\
&&(2c): \;- g^2 (-1+2\sin^2\theta_W) \sum_{k} \vert U_{k1} \vert^2~  
\hat C_{24}(M_Z^2;m_{\chi^+_k},m_{\tilde e_L}, m_{\tilde e_L}) \;, \nonumber\\
&&(2d): \;-{1\over 2} g^2  \sum_{k} \vert U_{k1} \vert^2~  
B_1(0;m_{\chi^+_k},m_{\tilde e_L})\;.
\end{eqnarray}
The resulting total corrections are 
\begin{eqnarray}
\delta h_{e_L}'&=& g^2 \biggl[ \sum_{ij} O_{ij}^{'L} V_{i1}^* V_{j1}\left[(2-d)\hat C_{24} 
+M_Z^2\hat C_{23}\right](M_Z^2;m_{\tilde \nu}, m_{\chi^+_i},m_{\chi^+_j}) \nonumber\\
&+& \sum_{ij} O_{ij}^{'R} V_{i1}^* V_{j1}~ m_{\chi^+_i} m_{\chi^+_j} 
\hat C_0(M_Z^2;m_{\tilde \nu}, m_{\chi^+_i},m_{\chi^+_j}) \nonumber\\
&-&
\sum_{k} \vert V_{k1} \vert^2~  
\hat C_{24}(M_Z^2;m_{\chi^+_k},m_{\tilde \nu}, m_{\tilde \nu})  \nonumber\\
&-&  \left(-{1\over2}+\sin^2\theta_W\right) \sum_{k} \vert V_{k1} \vert^2~  
B_1(0;m_{\chi^+_k},m_{\tilde \nu}) \biggr]\;,\\
\delta h_{\nu}' &=& -g^2 \biggl[ \sum_{ij} O_{ij}^{'R} U_{j1}^* U_{i1}\left[(2-d)\hat C_{24} 
+M_Z^2\hat C_{23}\right](M_Z^2;m_{\tilde e_L}, m_{\chi^+_j},m_{\chi^+_i}) \nonumber\\
&+&
\sum_{ij} O_{ij}^{'L} U_{j1}^* U_{i1}~ m_{\chi^+_i} m_{\chi^+_j} 
\hat C_0(M_Z^2;m_{\tilde e_L}, m_{\chi^+_j},m_{\chi^+_i})  \nonumber\\
&+&
(-1+2\sin^2\theta_W) \sum_{k} \vert U_{k1} \vert^2~  
\hat C_{24}(M_Z^2;m_{\chi^+_k},m_{\tilde e_L}, m_{\tilde e_L}) \nonumber\\
&+&
{1\over 2}   \sum_{k} \vert U_{k1} \vert^2~  
B_1(0;m_{\chi^+_k},m_{\tilde e_L}) \biggr]\;.
\end{eqnarray}
These corrections are finite as they should be. This can be seen from the 
relations
\begin{eqnarray}
&& \sum_{ij} O_{ij}^{'L} V_{i1}^* V_{j1}=
\sum_{ij} O_{ij}^{'R} U_{j1}^* U_{i1} = -1+\sin^2\theta_W \;, \nonumber\\
&& \sum_{i} V_{i1}^* V_{ik}=\sum_{i} U_{i1}^* U_{ik}=\delta_{1k}\;
\end{eqnarray}
and the fact ${\rm div}(\hat C_{24}) = -1/2~ {\rm div}(B_1)$ while $\hat C_0$ and
$\hat C_{23}$ are finite.

\subsection{Neutralino Contributions}

Because of their bino component, neutralinos induce corrections to 
both the left  and right  couplings of the leptons. 
Starting with the correction to the right-handed charged lepton coupling, we have
\begin{eqnarray}
&& \delta h_{e_R}'':\nonumber\\
&&(3a): \; -2g^2\tan^2\theta_W \sum_{ij} O_{ij}^{''L} N_{i1}^* N_{j1}\left[(2-d)\hat C_{24} 
+M_Z^2\hat C_{23}\right](M_Z^2;m_{\tilde e_R}, m_{\chi^0_j},m_{\chi^0_i}) \;, \nonumber\\
&&(3b): \; 2g^2\tan^2\theta_W \sum_{ij} O_{ij}^{''L} N_{j1}^* N_{i1}~ m_{\chi^0_i} m_{\chi^0_j}~ 
\hat C_0(M_Z^2;m_{\tilde e_R}, m_{\chi^0_i},m_{\chi^0_j}) \;, \nonumber\\
&&(3c): \;-4 g^2\tan^2\theta_W \sin^2\theta_W \sum_{k} \vert N_{k1} \vert^2~  
\hat C_{24}(M_Z^2;m_{\chi^0_k},m_{\tilde e_R}, m_{\tilde e_R}) \;, \nonumber\\
&&(3d): \;-2 g^2 \tan^2\theta_W \sin^2\theta_W  \sum_{k} \vert N_{k1} \vert^2~  
B_1(0;m_{\chi^0_k},m_{\tilde e_R})\;.
\end{eqnarray}
The corrections to the left-handed charged lepton coupling are given by
\begin{eqnarray}
&& \delta h_{e_L}'':\nonumber\\
&&(4a): \; -{g^2\over 2}\sum_{ij} O_{ij}^{''R} 
      (N_{j2}^*+\tan\theta_W N_{j1}^*) (N_{i2}+\tan\theta_W N_{i1})\nonumber\\
&&\;\;\;\;\;\;\;\;\;\;     \times \; \left[(2-d)\hat C_{24} 
       +M_Z^2\hat C_{23}\right](M_Z^2;m_{\tilde e_L}, m_{\chi^0_j},m_{\chi^0_i}) \;, \nonumber\\
&&(4b): \;{g^2\over 2} \sum_{ij} O_{ij}^{''R} 
(N_{i2}^*+\tan\theta_W N_{i1}^*) (N_{j2}+\tan\theta_W N_{j1})\nonumber\\
&&\;\;\;\;\;\;\;\;\;\;     \times \; m_{\chi^0_i} m_{\chi^0_j} ~
\hat C_0(M_Z^2;m_{\tilde e_L}, m_{\chi^0_i},m_{\chi^0_j}) \;, \nonumber\\
&&(4c): \;- g^2 \left( -{1\over 2}+\sin^2\theta_W \right)\sum_{k} 
       \vert N_{k2}+\tan\theta_W N_{k1} \vert^2~  
       \hat C_{24}(M_Z^2;m_{\chi^0_k},m_{\tilde e_L}, m_{\tilde e_L}) \;, \nonumber\\
&&(4d): \;- {g^2\over 2}\left( -{1\over 2}+\sin^2\theta_W \right)   \sum_{k} 
        \vert N_{k2}+\tan\theta_W N_{k1} \vert^2~  
        B_1(0;m_{\chi^0_k},m_{\tilde e_L})\;.
\end{eqnarray}
Finally, the neutrino coupling corrections are
\begin{eqnarray}
&& \delta h_{\nu}'':\nonumber\\
&&(5a): \; -{g^2\over 2}\sum_{ij} O_{ij}^{''R} 
      (N_{j2}^*-\tan\theta_W N_{j1}^*) (N_{i2}-\tan\theta_W N_{i1})\nonumber\\
&&\;\;\;\;\;\;\;\;\;\;     \times \; \left[(2-d)\hat C_{24} 
       +M_Z^2\hat C_{23}\right](M_Z^2;m_{\tilde \nu}, m_{\chi^0_j},m_{\chi^0_i}) \;, \nonumber\\
&&(5b): \;{g^2\over 2} \sum_{ij} O_{ij}^{''R} 
(N_{i2}^*-\tan\theta_W N_{i1}^*) (N_{j2}-\tan\theta_W N_{j1})\nonumber\\
&&\;\;\;\;\;\;\;\;\;\;     \times \; m_{\chi^0_i} m_{\chi^0_j} ~
\hat C_0(M_Z^2;m_{\tilde \nu}, m_{\chi^0_i},m_{\chi^0_j}) \;, \nonumber\\
&&(5c): \;- {g^2\over 2} \sum_{k} 
       \vert N_{k2}-\tan\theta_W N_{k1} \vert^2~  
       \hat C_{24}(M_Z^2;m_{\chi^0_k},m_{\tilde \nu}, m_{\tilde \nu}) \;, \nonumber\\
&&(5d): \;- {g^2\over 4}   \sum_{k} 
        \vert N_{k2}-\tan\theta_W N_{k1} \vert^2~  
        B_1(0;m_{\chi^0_k},m_{\tilde \nu})\;.
\end{eqnarray}
The total corrections are given by 
\begin{eqnarray}
\delta h_{e_R}'' &=& -2g^2 \tan^2\theta_W \biggl[
\sum_{ij} O_{ij}^{''L} N_{i1}^* N_{j1}\left[(2-d)\hat C_{24} 
+M_Z^2\hat C_{23}\right](M_Z^2;m_{\tilde e_R}, m_{\chi^0_j},m_{\chi^0_i})\nonumber\\
&-&
\sum_{ij} O_{ij}^{''L} N_{j1}^* N_{i1}~ m_{\chi^0_i} m_{\chi^0_j}~ 
\hat C_0(M_Z^2;m_{\tilde e_R}, m_{\chi^0_i},m_{\chi^0_j})\nonumber\\
&+&
 \sin^2 \theta_W \sum_{k} \vert N_{k1} \vert^2 \left\{  
2 \hat C_{24}(M_Z^2;m_{\chi^0_k},m_{\tilde e_R}, m_{\tilde e_R}) 
+  B_1(0;m_{\chi^0_k},m_{\tilde e_R}) \right\}
\biggr]\;.\\
\delta h_{e_L}'' &=&- {g^2\over 2} \biggl[
\sum_{ij} O_{ij}^{''R} 
      (N_{j2}^*+\tan\theta_W N_{j1}^*) (N_{i2}+\tan\theta_W N_{i1})\nonumber\\
&\times& \; \left[(2-d)\hat C_{24} 
       +M_Z^2\hat C_{23}\right](M_Z^2;m_{\tilde e_L}, m_{\chi^0_j},m_{\chi^0_i}) \nonumber\\
&-& \sum_{ij} O_{ij}^{''R} 
(N_{i2}^*+\tan\theta_W N_{i1}^*) (N_{j2}+\tan\theta_W N_{j1})\nonumber\\
&\times& \; m_{\chi^0_i} m_{\chi^0_j} ~
\hat C_0(M_Z^2;m_{\tilde e_L}, m_{\chi^0_i},m_{\chi^0_j}) +
\left( -{1\over2} + \sin^2\theta_W \right)
\sum_{k}  \vert N_{k2}+\tan\theta_W N_{k1} \vert^2 \nonumber\\
&\times& \left\{ 2 \hat C_{24}(M_Z^2;m_{\chi^0_k},m_{\tilde e_L}, m_{\tilde e_L}) +
 B_1(0;m_{\chi^0_k},m_{\tilde e_L}) \right\} 
\biggr].\\
\delta h_{\nu}'' &=& - {g^2\over 2} \biggl[ 
\sum_{ij} O_{ij}^{''R} 
      (N_{j2}^*-\tan\theta_W N_{j1}^*) (N_{i2}-\tan\theta_W N_{i1})\nonumber\\
&\times& \; \left[(2-d)\hat C_{24} 
       +M_Z^2\hat C_{23}\right](M_Z^2;m_{\tilde \nu}, m_{\chi^0_j},m_{\chi^0_i})\nonumber\\
&-& \sum_{ij} O_{ij}^{''R} 
(N_{i2}^*-\tan\theta_W N_{i1}^*) (N_{j2}-\tan\theta_W N_{j1})
 ~ m_{\chi^0_i} m_{\chi^0_j} ~
\hat C_0(M_Z^2;m_{\tilde \nu}, m_{\chi^0_i},m_{\chi^0_j}) \nonumber\\
&+& {1\over2} \sum_{k} 
       \vert N_{k2}-\tan\theta_W N_{k1} \vert^2 \left\{  
       2 \hat C_{24}(M_Z^2;m_{\chi^0_k},m_{\tilde \nu}, m_{\tilde \nu})
+ B_1(0;m_{\chi^0_k},m_{\tilde \nu}) \right\}
\biggr].
\end{eqnarray}
These expressions are finite due to the relations
\begin{eqnarray}
&& \sum_{ij} O_{ij}^{''L}N_{ik}^* N_{jl}=0 \;\;, \;\; (k,l= 1,2)\;, \nonumber\\
&& \sum_{ij} O_{ij}^{''R}N_{jk}^* N_{il}=0 \;
\end{eqnarray}
and the fact that the combination $2\hat C_{24}+B_1$ is finite. Note that the diagrams
in Figs. 3a, 4a, 5a are individually finite. The reason is transparent
in the weak eigenstates basis:  only the higgsinos couple to Z, and we 
retain only the gaugino coupling to the leptons, so a mass 
insertion is 
necessary on each  fermion line to complete the diagram.

\subsection{Decoupling of Heavy Superpartners}

In this subsection we demonstrate explicitly the decoupling of heavy SUSY particles.
As the SUSY mass scale increases, the gauginos and higgsinos become approximate mass
eigenstates and $V,U$ can be chosen such that 
\begin{eqnarray}
&& V_{ij} \rightarrow \delta_{ij} +{\cal O} \left({M_Z \over m_{susy}}\right)\;,\; 
U_{ij} \rightarrow \delta_{ij} + {\cal O}\left({M_Z \over m_{susy}}\right)\;,\nonumber\\
&& O^{'L}_{ij} \rightarrow \left(-1+\sin^2\theta_W \right)\delta_{i1}\delta_{j1}+
                         \left(-{1\over2}+\sin^2\theta_W \right)\delta_{i2}\delta_{j2}
+{\cal O}\left({M_Z\over m_{susy}}\right)\;,\nonumber\\
&& O^{'R}_{ij} \rightarrow \left(-1+\sin^2\theta_W \right)\delta_{i1}\delta_{j1}+
                         \left(-{1\over2}+\sin^2\theta_W\right)\delta_{i2}\delta_{j2}
+ {\cal O}\left({ M_Z\over m_{susy}}\right) \;.
\label{d}
\end{eqnarray}
In the expressions for the vertex structures $O^{'}_{ij}$, the factors in front
of the Kronecker delta symbols represent   the gaugino and higgsino couplings
to the Z boson.
It is clear that $O^{'}_{11}$ corresponds to the gaugino ($\tilde W^-$)
Z coupling, while $O^{'}_{22}$ corresponds to that of the higgsino 
($\tilde h^-$). Since $O^{'}_{ij}$ is to be contracted with $V_{i1}^*V_{j1}$,
the higgsino component drops out of all expressions in the decoupling
limit, as expected.
Denoting by $m$ a heavy scalar mass and by $M$ a heavy fermion mass,
we can rewrite $\delta h_{e_L}'$ as 
\begin{eqnarray}
\delta h_{e_L}'&=&g^2 \sin^2 \theta_W \biggl\{ \left[ (2-d)\hat C_{24} 
+M_Z^2\hat C_{23} \right] (0;m,M,M)  
+ M^2 \hat C_0(0;m,M,M) \nonumber\\
 &-& B_1(0;M,m) \biggr\} ~-~
g^2 \biggl\{ \left[ (2-d)\hat C_{24}  +M_Z^2\hat C_{23} \right] (0;m,M,M) \nonumber\\
&+& M^2 \hat C_0(0;m,M,M) +\hat C_{24}(0;M,m,m) -{1\over 2} B_1(0;M,m)
\biggr\} + {\cal O}\left({ M_Z^2\over m_{susy}^2}\right)\nonumber\\
&\longrightarrow& 0\;.
\end{eqnarray}
Each of the expressions in the curly brackets vanishes,
 see Appendix B. Note that even though the corrections in Eq.\ref{d} are linear
in $ M_Z / m_{susy}$, the SUSY contributions decouple quadratically, as they should. 
Similarly, for the neutrino final state we have
\begin{eqnarray}
\delta h_{\nu}'&=&-g^2 \sin^2 \theta_W \biggl\{ \left[ (2-d)\hat C_{24} 
+M_Z^2\hat C_{23} \right] (0;m,M,M)  
+ M^2 \hat C_0(0;m,M,M) \nonumber\\
 &+& 2 \hat C_{24} (0;M,m,m) \biggr\} ~+~
g^2 \biggl\{ \left[ (2-d)\hat C_{24}  +M_Z^2\hat C_{23} \right] (0;m,M,M) \nonumber\\
&+& M^2 \hat C_0(0;m,M,M) +\hat C_{24}(0;M,m,m) -{1\over 2} B_1(0;M,m)
\biggr\} + {\cal O}\left({ M_Z^2\over m_{susy}^2}\right) \nonumber\\
&\longrightarrow& 0\;.
\end{eqnarray}
Concerning the neutralino contributions, let us first consider $\delta h_{e_R}''$.
Since the mixing between the gauginos and higgsinos vanishes in the decoupling 
limit,
we have 
\begin{eqnarray}
&& N_{i1} \rightarrow \delta_{i1} + {\cal O}\left({ M_Z\over m_{susy}}\right) \;,\nonumber\\
&& O^{''L}_{ij} \rightarrow {\cal O}\left({ M_Z\over m_{susy}}\right)  \;\;\;{\rm for}\;\;\; i,j\not= 3,4\;.
\end{eqnarray}
As a result, the combination $O^{''L}_{ij}N_{i1}^* N_{j1}$ vanishes in this limit. Therefore 
\begin{eqnarray}
&&\delta h_{e_R}''= -2 g^2 \tan^2\theta_W \sin^2\theta_W 
\biggl\{
2 \hat C_{24}(0;M,m,m) 
+  B_1(0;M,m)
\biggr\} +{\cal O}\left({ M_Z^2\over m_{susy}^2}\right) \rightarrow 0\;.\nonumber
\end{eqnarray}
Again, the combination in the curly brackets vanishes (see Appendix B). The same arguments are valid
for the neutralino corrections to the couplings of the left-handed leptons.

\section{Numerical Analysis}

To separate out the effect of vertex corrections, we pursue the strategy of
Refs.\cite{Lebedev:2000ze}-\cite{Chang:2001xy},\cite{Takeuchi:1994zh}. 
That is, we utilize only those 
observables which can be expressed as ratios of the weak couplings. The effect
of oblique corrections \cite{Peskin:1990zt} 
then either cancels in the ratios  or can be absorbed into
effective $\sin^2\theta_W$. In the fit, we  leave $\sin^2\theta_W$ as a free
parameter and parameterize the vertex corrections as $\delta h_\nu$,
$\delta h_{l_L}$, and $\delta h_{l_R}$. In addition, we retain $\alpha_s(M_Z)$ 
as a free parameter since its value is determined from $R_l$.
The fit value of $\delta (\sin^2\theta_W)$ is not used for constraining the model
due to its model dependence. Specifically, $\delta (\sin^2\theta_W)$ 
 depends on the Higgs, squark, etc. 
masses and thus is not particularly useful in our general analysis.

We impose the following (direct search) constraints on the SUSY spectrum \cite{susylimits}:
\begin{eqnarray}
&&m_{\tilde e} \geq 99~GeV\;, \nonumber\\
&&m_{\tilde \mu} \geq 96~GeV\;, \nonumber\\
&&m_{\tilde \tau} \geq 87~GeV\;, \nonumber\\
&&m_{\tilde \nu} \geq 43~GeV\;, \nonumber\\
&&m_{ \chi^0} \geq 36~GeV\;, \nonumber\\
&&m_{ \chi^+} \geq 94~GeV\;.
\end{eqnarray}
We assume that the lepton parameters are generation-independent since 
lepton-universality breaking corrections are quite constrained 
(see, for example, the second reference in \cite{Lebedev:2000ze});
in any case this assumption is not important for our analysis. 

Before we proceed, a few comments are in order. First, note that $\mu$ is
determined by a particular combination of the Higgs mass parameters
(i.e. $m^2_{H_1}-m^2_{H_2} \tan^2\beta $  ). Thus having fixed $\mu$,
 the squark masses and the ``orthogonal'' combination of the Higgs mass parameters
remain free. This freedom results in the  uncertainty in the oblique corrections
mentioned above. However, to be specific, we will fix them in our numerical analysis
still assuming the freedom in the oblique corrections. Second, our results are presented
in terms of the high energy parameters. Since we do not generally assume a particular
framework or relations among the soft breaking parameters, one might wonder why not
interpret our results directly in terms of the low energy quantities.
However, not every low energy set of parameters can result from some high energy 
boundary conditions, especially consistent with radiative electroweak symmetry breaking.
To cite just one example, heavy gluinos and light squarks at low energies
are inconsistent with  high energy boundary conditions if we are
to avoid color breaking minima \cite{Dedes:2001nv}.
Thus, to be safe, we will generate each low energy set of parameters
via the RG running.

Numerically $R_{\nu/e}$  is   sensitive to the vertex corrections 
$\delta h_{i}$ and much less sensitive to the oblique corrections:
\begin{equation}
\delta R_{\nu/e}= 7.96 ~\delta h_{\nu} + 8.50~ \delta h_{e_L} - 7.33~  \delta h_{e_R}
 + 1.17 ~\delta s^2 \;,
\label{Rformula}
\end{equation}
where $s^2 \equiv \sin^2\theta_W$. 
Similarly, for the left-right asymmetries we have 
\begin{equation}
\delta A_{e}= -3.64~ \delta h_{e_L} - 4.23~  \delta h_{e_R}
 -7.87 ~\delta s^2 \;.
\label{Aformula}
\end{equation}
Since the SM prediction for $R_{\nu/e}$ is above the measured value whereas that for
the lepton asymmetries is below the measured values, $\delta h_{e_L}<0$ is 
favored
by both $R_{\nu/e}$ and $A_i$, $A_{FB}$. As shown below, $\delta h_{e_L}$ in the 
MSSM
is typically larger than $\delta h_{\nu}$ and $\delta h_{e_R}$. To get a feeling for
the value for $\delta h_{e_L}$ preferred by the fit, set 
$\delta h_{\nu}=\delta h_{e_R}=0$ and fit $R_i,A_i,A_{FB}$ with three parameters:
$\delta h_{e_L}$, $\delta \alpha_s$, and $\delta s^2$.
The best-fit values are
\begin{eqnarray}
&& \delta h_{e_L} = -0.00165 \pm 0.00096 \;,\nonumber\\
&& \delta \alpha_s = 0.024 \pm 0.014 \;,\nonumber\\
&& \delta s^2 = -0.0002 \pm  0.0005.
\end{eqnarray}
$R_{\nu/e}$ strongly pulls $\delta h_{e_L}$ to be negative, resulting 
in a large correction to $R_l$ which is in turn compensated by a 
large $\delta \alpha_s$. 
In addition to a genuine shift in $\alpha_s$, our ``effective''  $\delta \alpha_s$ 
parametrizes  potential corrections to $\Gamma(Z\rightarrow {\rm hadrons})$ from 
the squark/Higgs
sectors. This is, of course, just a ``toy'' fit. As we will see below, 
for a viable MSSM model
all the shifts are much smaller. Qualitatively, however,  the picture 
remains the same -- a negative   
$\delta h_{e_L}$ is preferred by the data.

\begin{table}
\begin{center}
\begin{tabular}{|c||c|c|c|c|c|}
\hline
 & $M_2$ = 150 & $M_2$ = 170 & $M_2$ = 190 & $M_2$ = 210 \\
\hline
\hline
$\chi^2$ & 11.09   & 11.30 & 11.40 & 11.48 \\
$\delta \alpha_{s} \times 10^{3}$ & $3.9\pm 3.9$ &$3.4\pm 3.9$  
& $3.1\pm 3.9$ & $2.8\pm 3.9$ \\
$\delta s^2 \times 10^4$ & $-9.9\pm 2.0$  & $-10.1\pm 2.0$ & 
$-10.2\pm 2.0$ & $-10.3\pm 2.0$ \\
\hline
\end{tabular}
\end{center}
\caption{
The quality of the fit $\chi^2$ as a function of the GUT scale parameter $M_2$ (GeV).
For comparison, the Standard Model gives  $\chi^2=11.44$ (d.o.f.=12-2).
The corresponding fit values of $\alpha_{s}$ and $\sin^2\theta_W$ are also displayed.
The other GUT scale parameters are  $m_{\tilde l}=10$ GeV,  $m_{\tilde e}=85$ GeV,
$M_1=100$ GeV,   $M_3=150$ GeV, $A=100$ GeV, $\tan\beta=5$ and the scalar mass parameter
(except for sleptons) is set to 100 GeV. }
\end{table}

Consider now the lepton vertex corrections in the MSSM.
In Figs.\ref{coupling}-\ref{coupling3} we display the vertex corrections 
$\delta h_\nu$, $\delta h_{e_L}$, and $\delta h_{e_R}$ as functions of
$M_2$ and $\tan\beta$.  Note that $\delta h_\nu$ and  $\delta h_{e_L}$
are quite sensitive to $M_2$, whereas its effect on $\delta h_{e_R}$ 
is negligible as it arises only via the RG running. In most of the parameter 
space, $\delta h_{e_L}$ dominates the other corrections; it has the right
sign (negative)  to mitigate
the invisible width ``anomaly'', especially for the positive sign of the 
$\mu$-term (which is also preferred by the $g_\mu-2$ measurement). 
We find that the regions of the parameter space
where SUSY contributions improve  the  agreement with the measured values of 
$g_\mu-2$ and $R_{\nu/e}$  
are generally compatible, see for instance Ref.\cite{Feng:2001tr}.

In Figs. \ref{R1} and \ref{R2}, we display
the corresponding shifts in $R_{\nu/e}$ as functions of $M_2$ and $\tan\beta$
(keeping $\sin^2 \theta_W$ fixed).  Varying $M_2$ from 135 to 250 GeV
corresponds to varying the light chargino mass from 95 to 180 GeV.
Fig. \ref{R4} shows the dependence of $R_{\nu/e}$ on the GUT left slepton mass
parameter $m_{\tilde l}$; its range 10-200 GeV translates into the slepton
mass $m_{\tilde e_L}$ range of 104-225 GeV.

The error bar for $R_{\nu/e}$
is 0.008 (see Table 2), so the supersymmetric contributions can only be responsible
for the shift of about $0.2\sigma$. This suppression results partly from the
cancellation between the neutrino and left-handed electron contributions.
Indeed, if the chargino is a pure gaugino and the sneutrino and left selectron masses
are equal,  $\delta h_\nu=-\delta h_{e_L}$ in the $M_Z \rightarrow 0$ 
approximation
and the resulting contribution to 
$R_{\nu/e}$ is very small (Eq.\ref{Rformula}).  Alternately, if the chargino
is a pure higgsino, the corresponding couplings are very much suppressed and
the resulting $\delta R_{\nu/e}$ is negligible.  One thus expects the largest
correction when there is a large splitting between the sneutrino and left selectron
masses (which is severely bounded by the SU(2) symmetry) and/or when 
the chargino is a  gaugino-higgsino mixture ($M_2/\mu \sim 1$)
\footnote{This was also noted in Ref.\cite{kaoru}}.
  
The dependence on other input parameters is significantly weaker. An increase in
$M_3$ affects  the $\mu$-term via the radiative EW symmetry breaking condition, which in turn
results in heavier charginos and neutralinos. The effect of $M_1$ is not
significant due to the subdominant role of the neutralino contributions.
For the same reason the dependence on the masses of the right sleptons is weak.

For completeness, below we provide representative low energy parameters for our studies.
The GUT scale parameters  $m_{\tilde l}=10$ GeV,  $m_{\tilde e}=85$ GeV, $m_0=100$ GeV,
$M_1=100$ GeV, $M_2=135$ GeV, $M_3=200$ GeV, $A=100$ GeV, $\phi_\mu=0$,
where $m_0$
is the mass parameter for the scalars other than sleptons,
lead to the following low energy spectrum
\begin{eqnarray}
&& m_{\chi^+_i} \simeq (372,95)\;GeV \;, \;
   m_{\chi^0_i} \simeq (374,353,96,38)\;GeV \;, \nonumber\\
&& m_{\tilde \nu} \simeq 75\;GeV\;,\; m_{\tilde e_L} \simeq 104\;GeV\;,\; 
   m_{\tilde e_R} \simeq 101\;GeV\;,\; \nonumber 
\end{eqnarray}
and the following mixing matrices
\begin{eqnarray}
&& U= \left( \matrix{0.34 & 0.94 \cr
                    0.94 & -0.34 } \right) \;,\;
   V= \left( \matrix{0.19 & 0.98 \cr
                    0.98 & -0.19 } \right) \;,\; \nonumber\\
&& N= \left( \matrix{ 0.11 & -0.26 & -0.67 & 0.69  \cr
                      0.05{\rm i} & -0.08{\rm i} & -0.70 {\rm i}& -0.71 {\rm i}\cr
                      0.15 & 0.95 & -0.22 & 0.12 \cr
                     -0.98 & 0.11 & -0.15 & 0.06  } \right) \;.
\end{eqnarray}

We now turn to the discussion of the fit.
In Table 1 we present our fit results for different values of $M_2$.
That is, we fix the lepton vertex corrections using our GUT parameters  
in the fit
and calculate the corresponding $\chi^2$ (d.o.f.= 12-2).   
The parameters $\delta s^2$ and (effective) $\delta \alpha_s$ are left
as free fit parameters, 
which means that the Higgs and the squark sectors 
``adjust'' themselves so as to give the best fit results.  

To determine if there is any improvement over the SM, 
we perform a fit for the SM under the {\it same circumstances}, 
i.e. lepton vertex corrections set to zero, $\delta s^2$ and 
$\delta \alpha_s$ free to account for a variation in the Higgs mass and $\alpha_s$. 
The Standard Model fit gives 
\begin{eqnarray}
&& \chi^2=11.44 \;\; ({\rm d.o.f.}= 12-2) \;, \nonumber\\
&& \delta \alpha_s = 0.0020 \pm 0.0039\;,\nonumber\\
&& \delta s^2 =-0.00103 \pm 0.00020 \;.
\end{eqnarray}
If the chargino is light (100 GeV), the MSSM fit gives $\chi^2=11.09$.
We see that the SUSY vertex corrections indeed improve the fit 
due to the improvement in $R_{\nu/ e}$ and the lepton asymmetries.
The quality of 
the fit quickly approaches that of the SM as the chargino mass increases. 
We note that the best fit value of $\delta s^2$ for both the SM and the MSSM 
significantly deviates from zero because of the SLD asymmetries, which
signifies that the light Higgs is preferred.

 \section{Conclusions}

We have analyzed leptonic electroweak observables in the general MSSM.
We find that supersymmetry can mitigate some of the discrepancies between
the Standard Model predictions and the observed values. Namely,
it produces vertex corrections of the right sign to improve agreement 
with $R_{\nu/e}$ and the leptonic asymmetries. As a result, the 
electroweak fit is improved from $\chi^2=11.44$ (SM) 
to $\chi^2=11.09$ (MSSM). This required a light ($\sim 100$ GeV)
chargino and relatively light  (100-250 GeV) sleptons.

Although the improvement from statistical point of view is not
very significant, it is quite encouraging since in the same region of
the parameter space the $g_\mu-2$  discrepancy is also mitigated.
This is to be contrasted with a number of ``new physics'' models considered
earlier \cite{Lebedev:2000ze}-\cite{Chang:2001xy}, all of which made the
electroweak fit worse. The improvement of the fit requires light superpartners
which can be  detected in collider experiments in the near future.

{\bf Acknowledgements.} The authors are indebted to G. Kane and T. Takeuchi
for very helpful discussions and suggestions.

\appendix
\section{Appendix: Notation and Conventions}

We use the following (chiral) representation of the Dirac matrices:
\begin{eqnarray}
&& \gamma_{\mu}=\left( \matrix{0&\sigma_{\mu}\cr
                          \bar \sigma_{\mu}&0}\right)\;,\;
\gamma_5=i\gamma^0\gamma^1\gamma^2 \gamma^3=
\left( \matrix{-{\bf 1}&0\cr
                          0&{\bf 1}}\right)\;,\;
\end{eqnarray}
where $\sigma^\mu=(1,{\overrightarrow \sigma})$ and $\bar \sigma^\mu=(1,-{\overrightarrow \sigma})$.
The  left and right components of a Dirac spinor and the corresponding projectors
are defined by
\begin{eqnarray}
&& \psi\equiv \left( \matrix{\psi_L\cr
                             \psi_R} \right) \;,\;
P_{L,R}\equiv {1\over 2}(1\mp\gamma_5)\;.
\end{eqnarray}
The charge conjugated spinor is given by
\begin{equation}
\psi^c=C\bar\psi^T \;,\; C=-i\gamma^2\gamma^0 \;,\; \bar\psi\equiv\psi^\dagger \gamma^0 \;.
\end{equation}
In terms of the two-component spinors this corresponds to
\begin{equation}
\psi^c = \left( \matrix{-i\sigma^2 \psi_R^*\cr
                             i\sigma^2\psi_L^*} \right) \;,\;
{\overline {\psi^c}}= \left(\psi_L^T (-i\sigma^2), \psi_R^T (i\sigma^2)\right)\;.
\end{equation}
Free fermions satisfy the following Dirac equation in the two-component notation
\begin{eqnarray}
&& (k\cdot \bar \sigma)~ \psi_L =m~\psi_R \;,\nonumber\\
&& (k\cdot  \sigma)~ \psi_R =m~\psi_L \;,
\end{eqnarray}
The corresponding propagators read
\begin{eqnarray}
&&\langle \psi_L ~\psi_L^\dagger\rangle=i {k\cdot \sigma \over k^2-m^2} \;\;,\;\;  
\langle \psi_R ~\psi_R^\dagger\rangle=i {k\cdot \bar\sigma \over k^2-m^2} \;, \nonumber\\
&&\langle \psi_L ~\psi_R^\dagger\rangle=\langle \psi_R ~\psi_L^\dagger\rangle=
i {m \over k^2-m^2}\;.
\end{eqnarray}
The following identities are useful for calculating Feynman diagrams in terms of
the two-component spinors:
\begin{eqnarray}
&& \sigma^2~ \sigma_\mu^T ~\sigma^2 = \bar \sigma_\mu \;\;, \;\;
   \sigma^2 ~\bar\sigma_\mu^T ~\sigma^2 =  \sigma_\mu \;, \; \nonumber\\
&& \bar \sigma^\nu ~\sigma^\mu ~ \bar \sigma_\nu = (2-d)~\bar \sigma^\mu \;,\nonumber\\
&& (p \cdot \bar \sigma)~ \sigma^\mu ~(p \cdot \bar \sigma) = 
-p^2 ~\bar \sigma^\mu \;.
\end{eqnarray}

\section{Feynman Integrals}

\newcommand{\pole}{\Delta_\epsilon}

\setlength{\jot}{2mm}
\setlength{\abovedisplayskip}{4mm plus 1mm minus 1mm}
\setlength{\belowdisplayskip}{4mm plus 1mm minus 1mm}

Here we make explicit our notation for the scalar and tensor integrals 
that appear in the calculation.
The definitions of the integrals are slightly different from  those of
Ref.~\cite{'tHooft:1979xw}. 
The hat on the tensor integrals serves as a reminder of these differences.

\subsection{Scalar Integrals}

\noindent
We define the functions $B_0$ and $\hat{C}_0$ by:
\begin{eqnarray}
B_0\,[\,p^2;m_1,m_2\,] & \equiv & 
i\,\mu^{4-d}\int\!\frac{ d^d k }{ (2\pi)^d }\, 
\frac{ 1 }{ (k^2 - m_1^2)\,[\,(k+p)^2 - m_2^2\,] }, \\
\hat{C}_0\,[\,p^2,q^2,(p-q)^2;m_1,m_2,m_3\,] & \equiv &
i\int\!\frac{ d^4 k }{ (2\pi)^4 }\, 
\frac{ 1 }
     { ( k^2 - m_1^2 )\,[\,(k+p)^2 - m_2^2\,]\,[\,(k+q)^2 - m_3^2\,] }.\nonumber
\end{eqnarray}
The general form of $B_0$ is given by
\begin{displaymath}
B_0\,[\,p^2;m_1,m_2\,] 
= \frac{-1}{(4\pi)^2}
      \left[ \pole 
             - \frac{ m_1^2\ln(m_1^2/\mu^2) - m_2^2\ln(m_2^2/\mu^2) }
                     { m_1^2 - m_2^2 }
             + 1 + F(p^2;m_1,m_2)
      \right],
\end{displaymath}
where $\pole = \frac{2}{4-d} - \gamma_E + \ln{4\pi}$, and \cite{HOLLIK:90}
\begin{eqnarray}
F(p^2;m_1,m_2) &=& 1
  + \frac{1}{2} \left( \frac{\Sigma}{\Delta} - \Delta 
                \right) \ln\left( \frac{m_1^2}{m_2^2}
                           \right) \nonumber\\
 & -& \frac{1}{2}\sqrt{ 1 - 2\Sigma + \Delta^2 } 
  \,\ln\left( \frac{ 1 - \Sigma + \sqrt{ 1 - 2\Sigma + \Delta^2 } }
                    { 1 - \Sigma - \sqrt{ 1 - 2\Sigma + \Delta^2 } }
       \right)
\end{eqnarray}
with
\begin{equation}
\Sigma \equiv \frac{m_1^2 + m_2^2}{p^2},\;\;
\Delta \equiv \frac{m_1^2 - m_2^2}{p^2}.
\label{SigmaDelta}
\end{equation}
The function $F(p^2;m_1,m_2)$ vanishes in the limit
$p^2\rightarrow 0$.   
The general form of the $\hat{C}_0$ function is fairly complex
and we refer the reader to Ref.~\cite{'tHooft:1979xw}. The 
special case relevant for our calculations is 
\begin{eqnarray}
\hat{C}_0\,[\,0,0,Q^2;m_1,m_2,m_3\,]
& = & \frac{ 1 }{ (4\pi)^2 } \int_0^1\!dy\,
      \frac{ 1 }{ m_3^2-m_1^2-yQ^2 } \nonumber\\
& \times &  \ln\left[ \frac{y(y-1)Q^2+(m_2^2-m_3^2)y +m_3^2 }{ y(m_2^2-m_1^2)+m_1^2 }
         \right]. 
\end{eqnarray}

\subsection{Tensor Integrals}

\noindent
Definition and general form of $B_1$:
\begin{eqnarray}
B_\mu\,[\,p;m_1,m_2\,]
& = & i\mu^{4-d} \int\!\frac{ d^dk }{ (2\pi)^d }\,
      \frac{ k_\mu }{ (k^2 - m_1^2)\,[(k+p)^2 - m_2^2] }
\;\equiv\; p_{\mu} B_1\,[\,p^2;m_1,m_2\,], \nonumber \\ 
B_1\,[\,p^2;m_1,m_2\,]
& = & - \frac{1}{2} B_0\,[\,p^2;m_1,m_2\,]
      + \frac{ 1 }{ (4\pi)^2 }
        \left( \frac{ m_1^2 - m_2^2 }{ 2p^2 }
        \right)
        F(p^2;m_1,m_2).
\end{eqnarray}
Note the following useful relations among the $B$--functions:
\begin{eqnarray}
0 & = & B_0\,[\,p^2;m_1,m_2\,]
      + B_1\,[\,p^2;m_1,m_2\,]
      + B_1\,[\,p^2;m_2,m_1\,],\\ 
0 & = & (m_1^2-m_2^2)\,B_0\,[\,0;m_1,m_2\,]
      + (m_2^2-m_3^2)\,B_0\,[\,0;m_2,m_3\,]
      + (m_3^2-m_1^2)\,B_0\,[\,0;m_3,m_1\,].\nonumber
\end{eqnarray}
Definition of the $C$--functions: (Note the difference from the
definitions in Ref.~\cite{'tHooft:1979xw}.)
\begin{eqnarray}
C_\mu\,[\,p,q;m_1,m_2,m_3\,]
& = & i\int\!\frac{ d^4 k }{ (2\pi)^4 }
      \frac{ k_\mu }
           { (k^2 - m_1^2)\,[ (k+p)^2 - m_2^2 ]\,[ (k+q)^2 - m_3^2 ] }
\nonumber\\
& \equiv & p_\mu \hat{C}_{11} + q_\mu \hat{C}_{12},
\\
C_{\mu\nu}[\,p,q;m_1,m_2,m_3\,]
& = & i\mu^{4-d}\int\!\frac{ d^d k }{ (2\pi)^d }
      \frac{ k_\mu k_\nu }
           { (k^2 - m_1^2)\,[ (k+p)^2 - m_2^2 ]\,[ (k+q)^2 - m_3^2 ] }
\nonumber\\
& \equiv & p_\mu p_\nu \hat{C}_{21} + q_\mu q_\nu \hat{C}_{22}
          + ( p_\mu q_\nu + q_\mu p_\nu ) \hat{C}_{23}
          + g_{\mu\nu} \hat{C}_{24}. \nonumber
\end{eqnarray}
For the purpose of this paper, we will only need to evaluate these functions
for $p^2 = q^2 = 0$ (we neglect final state fermion masses).
$Q^2 = (p-q)^2 = -2p\cdot q$ will then be the invariant mass squared of the
initial vector boson.  For this parameter choice, 
the $C$--functions can be expressed in terms of the $B$--functions
and $\hat{C}_0$ as:
\begin{eqnarray}
&&\hat{C}_{11}
 =  -\frac{ 1 }{ Q^2 }
       \left\{ B_0\,[\,  0;m_1,m_2\,]
             - B_0\,[\,Q^2;m_2,m_3\,] 
             - (m_1^2-m_3^2)\,
               \hat{C}_0
       \right\},\nonumber
\\
&&\hat{C}_{12}
 =  -\frac{ 1 }{ Q^2 }
       \left\{ B_0\,[\,  0;m_1,m_3\,]
             - B_0\,[\,Q^2;m_2,m_3\,] 
             - (m_1^2-m_2^2)\,
               \hat{C}_0
       \right\}, \nonumber
\\
&&\hat{C}_{24}
 = {1\over 2} \biggl[ - B_1\,[\,Q^2;m_2,m_3\,]
      + (m_1^2-m_2^2)\,\hat{C}_{11} + m_1^2\,\hat{C}_0 -{1\over 2(4\pi)^2}\biggr], \nonumber
\\
&&(d-2)\,\hat{C}_{24}-Q^2\,\hat{C}_{23}
 =  - B_1\,[\,Q^2;m_3,m_2\,] - (m_1^2-m_2^2)\,\hat{C}_{11}+{1\over 2(4\pi)^2}\;.
\end{eqnarray}
We do not list expressions for $\hat{C}_{21}$ nor $\hat{C}_{22}$
since we do not use them in this paper.

\subsection{Decoupling Limit}

Below we list approximate formulas valid in the decoupling limit $p^2/m^2_{s,f} \rightarrow
0$. Here
$m_s$ and $m_f$ denote the scalar and fermion masses, respectively.
Omitting the ${\cal O}(p^2/m^2_{s,f})$ terms, we have
\begin{eqnarray}
&&[(d-2)\,\hat{C}_{24}- p^2\,\hat{C}_{23}]\left( 0,0,p^2;m_{s},m_{f},m_{f}
                          \right) 
\approx -\frac{1}{(4\pi)^2} 
\left[ \frac{1}{2} \left( \pole - \ln{\frac{m_f^2}{\mu^2}} \right) + f(x)
\right], \nonumber\\
&&\hat{C}_{24}\left( 0,0,p^2;m_{f},m_{s},m_{s} \right)
\;\approx\; 
- \frac{1}{2(4\pi)^2}
\left[ \frac{1}{2} \left( \pole - \ln{\frac{m_f^2}{\mu^2}} \right) - g(x)
\right], 
\nonumber\\
&& m_f^2\,\hat{C}_{0}\left( 0,0,p^2;m_{s},m_{f},m_{f} \right) 
\;\approx\; 
-\frac{1}{(4\pi)^2} \left[\, f(x)+ g(x) \,\right],
\nonumber\\
&& \hat{B}_{1} \left( 0;m_{f},m_{s} \right)
\;\approx\;
\frac{1}{(4\pi)^2}
\left[ \frac{1}{2} \left( \pole - \ln{\frac{m_f^2}{\mu^2}} \right) - g(x)
\right],
\end{eqnarray}
where 
\begin{eqnarray}
f(x) & = & -\frac{1}{4 (1-x)^2}\left(\,x^2 - 1 - 2 \ln{x} \,\right), \cr
g(x) & = & -\frac{1}{2} \ln{x} + \frac{1}{4 (1-x)^2}
            \left[ -(1-x)(1-3 x) + 2 x^2 \ln{x} \,\right]
\end{eqnarray}
for $x = m_f^2 / m_s^2.$


\newpage
\begin{table}[p]
\begin{center}
\begin{tabular}{|c|c|c|}
\hline
Observable & Measured Value & ZFITTER Prediction \\
\hline\hline
\multicolumn{2}{|l|}{\underline{$Z$ lineshape variables}} & \\
$m_Z$                & $91.1876 \pm 0.0021$ GeV & input       \\
$\Gamma_Z$           & $2.4952 \pm 0.0023$ GeV  & unused      \\
$\sigma_{\rm had}^0$ & $41.541 \pm 0.037$ nb    & unused  \\
$R_e$                & $20.804 \pm 0.050$       & $20.739$ \\
$R_\mu$              & $20.785 \pm 0.033$       & $20.739$ \\
$R_\tau$             & $20.764 \pm 0.045$       & $20.786$ \\
$A_{\rm FB}(e   )$   & $0.0145 \pm 0.0025$      & $0.0152$ \\
$A_{\rm FB}(\mu )$   & $0.0169 \pm 0.0013$      & $0.0152$ \\
$A_{\rm FB}(\tau)$   & $0.0188 \pm 0.0017$      & $0.0152$ \\
$R_{\nu /e}      $   & $1.9755 \pm 0.0080$      & $1.9916$ \\
\hline
\multicolumn{2}{|l|}{\underline{$\tau$ polarization at LEP}} & \\
$A_e$        & $0.1498 \pm 0.0048$      & $0.1423$ \\ 
$A_\tau$     & $0.1439 \pm 0.0042$      & $0.1424$ \\
\hline
\multicolumn{2}{|l|}{\underline{SLD left--right asymmetries}} & \\
$A_{LR}$     & $0.1514 \pm 0.0022$    & $0.1423$ \\
$A_e$        & $0.1544 \pm 0.0060$    & $0.1423$ \\
$A_{\mu}$    & $0.142 \pm 0.015$    & $0.1423$ \\
$A_{\tau}$   & $0.136 \pm 0.015$    & $0.1424$ \\
\hline
\end{tabular}
\caption{LEP/SLD observables 
and their Standard Model predictions.
The data are from Refs.\cite{Swartz:2000xv} and \cite{LEP:99}.
The Standard Model predictions were calculated using ZFITTER v.6.21 
\cite{ZFITTER:99} with $m_t = 174.3$~GeV,
$m_H = 300$~GeV, and $\alpha_s(m_Z) = 0.120$ as input.}
\label{LEP-SLD-DATA}
\end{center}
\end{table}

\medskip

\begin{table}[ht]
\begin{center}
\begin{tabular}{|c|ccccccccc|}
\hline
& $m_Z$     & $\Gamma_Z$     & $\sigma_{\rm had}^0$
& $R_e$     & $R_\mu$     & $R_\tau$ 
& $A_{\rm FB}(e)$ & $A_{\rm FB}(\mu)$ & $A_{\rm FB}(\tau)$ \\
\hline
$m_Z$ 
& $1.000$            & $-0.008$           & $-0.050$ 
& $\phantom{-}0.073$ & $\phantom{-}0.001$ & $\phantom{-}0.002$ 
& $-0.015$           & $\phantom{-}0.046$ & $\phantom{-}0.034$ \\
$\Gamma_Z$
&                    & $\phantom{-}1.000$ & $-0.284$ 
& $-0.006$           & $\phantom{-}0.008$ & $\phantom{-}0.000$ 
& $-0.002$           & $\phantom{-}0.002$ & $-0.003$           \\
$\sigma_{\rm had}^0$
&                    &                    & $\phantom{-}1.000$ 
& $\phantom{-}0.109$ & $\phantom{-}0.137$ & $\phantom{-}0.100$ 
& $\phantom{-}0.008$ & $\phantom{-}0.001$ & $\phantom{-}0.007$ \\
$R_e$
&                    &                    & 
& $\phantom{-}1.000$ & $\phantom{-}0.070$ & $\phantom{-}0.044$ 
& $-0.356$           & $\phantom{-}0.023$ & $\phantom{-}0.016$ \\
$R_\mu$
&                    &                    & 
&                    & $\phantom{-}1.000$ & $\phantom{-}0.072$ 
& $\phantom{-}0.005$ & $\phantom{-}0.006$ & $\phantom{-}0.004$ \\
$R_\tau$ 
&                    &                    &
&                    &                    & $\phantom{-}1.000$ 
& $\phantom{-}0.003$ & $-0.003$           & $\phantom{-}0.010$ \\
$A_{\rm FB}(e)$ 
&                    &                    &
&                    &                    & 
& $\phantom{-}1.000$ & $-0.026$           & $-0.020$ \\
$A_{\rm FB}(\mu)$ 
&                    &                    & 
&                    &                    & 
&                    & $\phantom{-}1.000$ & $\phantom{-}0.045$ \\
$A_{\rm FB}(\tau)$
&                    &                    & 
&                    &                    & 
&                    &                    & $\phantom{-}1.000$ \\
\hline
\end{tabular}
\caption{The correlation of the $Z$ lineshape variables at LEP. The correlation
of $R_{\nu /e}$ with $A_{\rm FB}(e)$ is +0.28, while its correlation with the
$\mu$ and $\tau$ observables is  negligible. }
\label{LEPcorrelations}
\end{center}
\end{table}
\clearpage

\newpage

\begin{figure}[ht]
\begin{center}
\unitlength=1cm
\begin{picture}(17,6)(0,0)
\unitlength=1mm

\put(3,33){$Z$}
\put(23,42){${\tilde \chi}_i^+$}
\put(23,25){${\tilde \chi}_j^+$}
\put(37,33){${\tilde \nu}_L$}
\put(46,47){$e_L$}
\put(46,20){$e_L$}

\put(61,42){${\tilde \chi}_i^+$}
\put(61,25){${\tilde \chi}_j^+$}
\put(75,33){${\tilde \nu}_L$}
\put(84,47){$e_L$}
\put(84,20){$e_L$}

\put(101,42){${\tilde \nu}_L$}
\put(101,25){${\tilde \nu}_L$}
\put(113,33){${\tilde \chi}_i^+$}
\put(122,47){$e_L$}
\put(122,20){$e_L$}

\put(141,23){${\tilde \chi}_i^+$}
\put(151,33){${\tilde \nu}_L$}
\put(131,29){$e_L$}
\put(160,47){$e_L$}
\put(160,20){$e_L$}

\put(23.25,10){(a)}
\put(63.75,10){(b)}
\put(103.25,10){(c)}
\put(140.75,10){(d)}
\epsfbox[0 600 480 780]{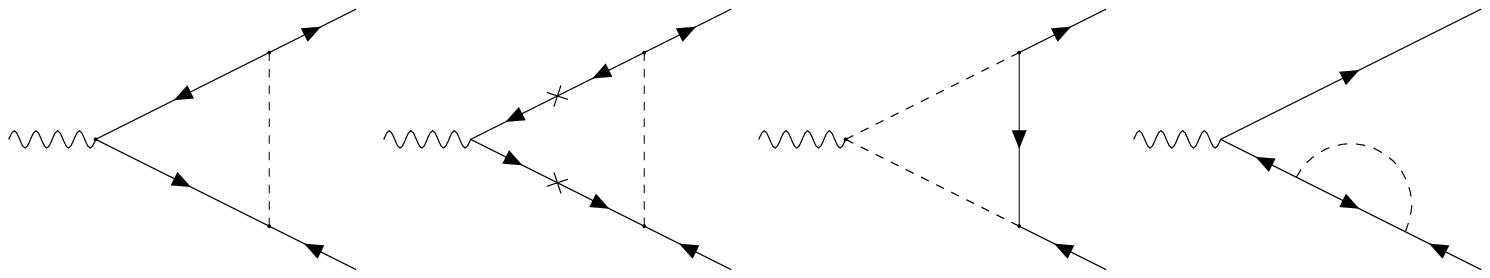}
\end{picture}
\caption{Chargino-sneutrino corrections to the $Z e_L {\overline e}_L$ 
vertex (in this and other figures the wave function renormalization diagram
with a loop on the upper fermion leg is not shown).} 
\label{FIG1}
\end{center}
\end{figure}

\medskip

\medskip

\begin{figure}[ht]
\begin{center}
\unitlength=1cm
\begin{picture}(17,6)(0,0)
\unitlength=1mm

\put(3,33){$Z$}
\put(23,42){${\tilde \chi}_i^+$}
\put(23,25){${\tilde \chi}_j^+$}
\put(37,33){${\tilde e}_L$}
\put(46,47){$\nu_L$}
\put(46,20){$\nu_L$}

\put(61,42){${\tilde \chi}_i^+$}
\put(61,25){${\tilde \chi}_j^+$}
\put(75,33){${\tilde e}_L$}
\put(84,47){$\nu_L$}
\put(84,20){$\nu_L$}

\put(101,42){${\tilde e}_L$}
\put(101,25){${\tilde e}_L$}
\put(113,33){${\tilde \chi}_i^+$}
\put(122,47){$\nu_L$}
\put(122,20){$\nu_L$}

\put(141,23){${\tilde \chi}_i^+$}
\put(151,33){${\tilde e}_L$}
\put(131,29){$\nu_L$}
\put(160,47){$\nu_L$}
\put(160,20){$\nu_L$}

\put(23.25,10){(a)}
\put(63.75,10){(b)}
\put(103.25,10){(c)}
\put(140.75,10){(d)}

\epsfbox[0 600 480 780]{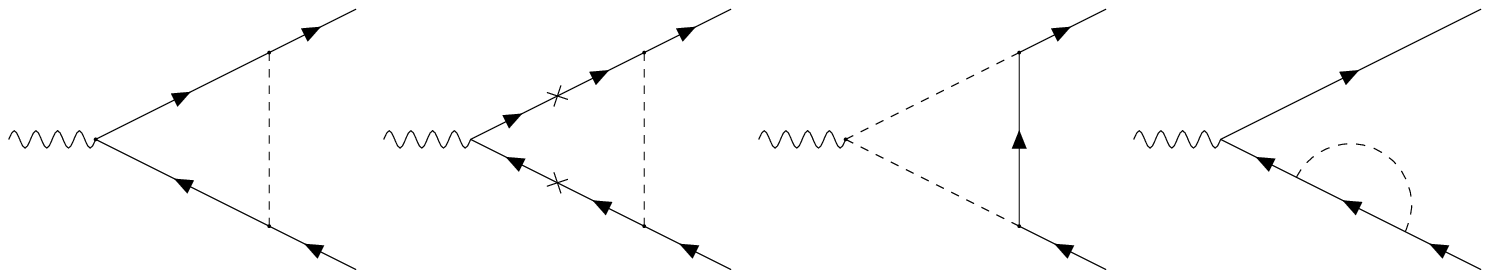}
\end{picture}
\caption{Chargino-selectron corrections to the $Z \nu_L {\overline \nu}_L$ vertex.} 
\label{FIG2}
\end{center}
\end{figure}

\begin{figure}[ht]
\begin{center}
\unitlength=1cm
\begin{picture}(17,6)(0,0)
\unitlength=1mm

\put(3,33){$Z$}
\put(23,42){${\tilde \chi}_i^0$}
\put(23,25){${\tilde \chi}_j^0$}
\put(37,33){${\tilde e}_R$}
\put(46,47){$e_R$}
\put(46,20){$e_R$}

\put(61,42){${\tilde \chi}_i^0$}
\put(61,25){${\tilde \chi}_j^0$}
\put(75,33){${\tilde e}_R$}
\put(84,47){$e_R$}
\put(84,20){$e_R$}

\put(101,42){${\tilde e}_R$}
\put(101,25){${\tilde e}_R$}
\put(113,33){${\tilde \chi}_i^0$}
\put(122,47){$e_R$}
\put(122,20){$e_R$}

\put(141,23){${\tilde \chi}_i^0$}
\put(151,33){${\tilde e}_R$}
\put(131,29){$e_R$}
\put(160,47){$e_R$}
\put(160,20){$e_R$}

\put(23.25,10){(a)}
\put(63.75,10){(b)}
\put(103.25,10){(c)}
\put(140.75,10){(d)}

\epsfbox[0 600 480 780]{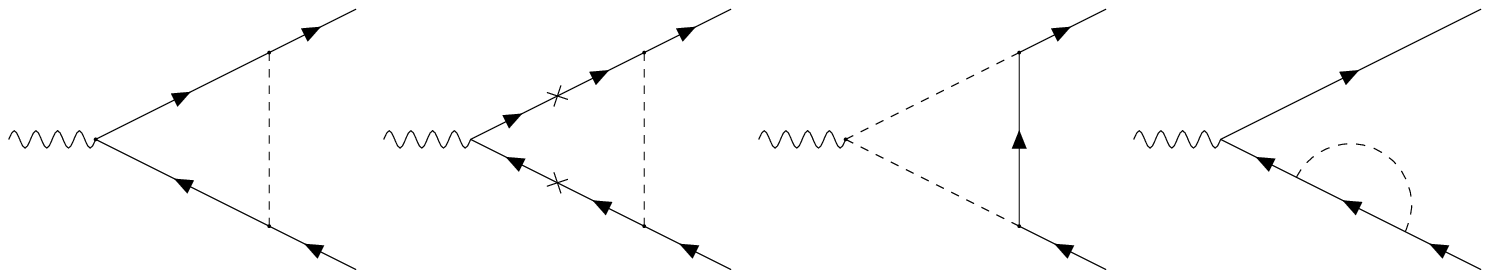}
\end{picture}
\caption{Neutralino-selectron corrections to the $Z e_R {\overline e}_R$ vertex.} 
\label{FIG3}
\end{center}
\end{figure}

\begin{figure}[ht]
\begin{center}
\unitlength=1cm
\begin{picture}(17,6)(0,0)
\unitlength=1mm

\put(3,33){$Z$}
\put(23,42){${\tilde \chi}_i^0$}
\put(23,25){${\tilde \chi}_j^0$}
\put(37,33){${\tilde e}_L$}
\put(46,47){$e_L$}
\put(46,20){$e_L$}

\put(61,42){${\tilde \chi}_i^0$}
\put(61,25){${\tilde \chi}_j^0$}
\put(75,33){${\tilde e}_L$}
\put(84,47){$e_L$}
\put(84,20){$e_L$}

\put(101,42){${\tilde e}_L$}
\put(101,25){${\tilde e}_L$}
\put(113,33){${\tilde \chi}_i^0$}
\put(122,47){$e_L$}
\put(122,20){$e_L$}

\put(141,23){${\tilde \chi}_i^0$}
\put(151,33){${\tilde e}_L$}
\put(131,29){$e_L$}
\put(160,47){$e_L$}
\put(160,20){$e_L$}

\put(23.25,10){(a)}
\put(63.75,10){(b)}
\put(103.25,10){(c)}
\put(140.75,10){(d)}

\epsfbox[0 600 480 780]{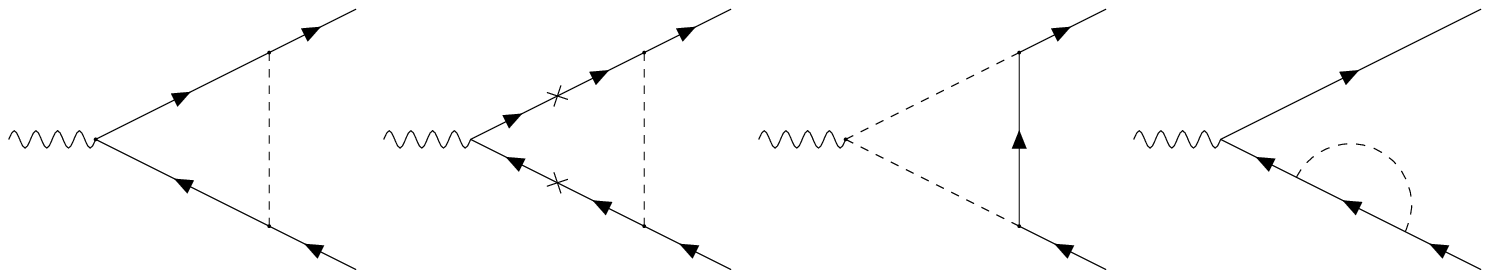}
\end{picture}
\caption{Neutralino-selectron corrections to the $Z e_L {\overline e}_L$ vertex.} 
\label{FIG4}
\end{center}
\end{figure}

\begin{figure}[ht]
\begin{center}
\unitlength=1cm
\begin{picture}(17,6)(0,0)
\unitlength=1mm

\put(3,33){$Z$}
\put(23,42){${\tilde \chi}_i^0$}
\put(23,25){${\tilde \chi}_j^0$}
\put(37,33){${\tilde \nu}_L$}
\put(46,47){$\nu_L$}
\put(46,20){$\nu_L$}

\put(61,42){${\tilde \chi}_i^0$}
\put(61,25){${\tilde \chi}_j^0$}
\put(75,33){${\tilde \nu}_L$}
\put(84,47){$\nu_L$}
\put(84,20){$\nu_L$}

\put(101,42){${\tilde \nu}_L$}
\put(101,25){${\tilde \nu}_L$}
\put(113,33){${\tilde \chi}_i^0$}
\put(122,47){$\nu_L$}
\put(122,20){$\nu_L$}

\put(141,23){${\tilde \chi}_i^0$}
\put(151,33){${\tilde \nu}_L$}
\put(131,29){$\nu_L$}
\put(160,47){$\nu_L$}
\put(160,20){$\nu_L$}

\put(23.25,10){(a)}
\put(63.75,10){(b)}
\put(103.25,10){(c)}
\put(140.75,10){(d)}

\epsfbox[0 600 480 780]{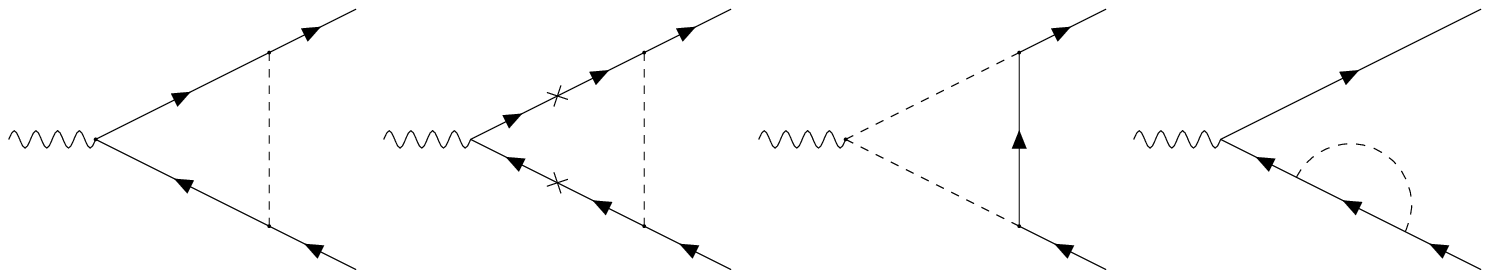}
\end{picture}
\caption{Neutralino-sneutrino corrections to the $Z \nu_L {\overline \nu}_L$ vertex.} 
\label{FIG5}
\end{center}
\end{figure}

\begin{figure}[ht]
\begin{center}
\begin{tabular}{c}
\epsfig{file=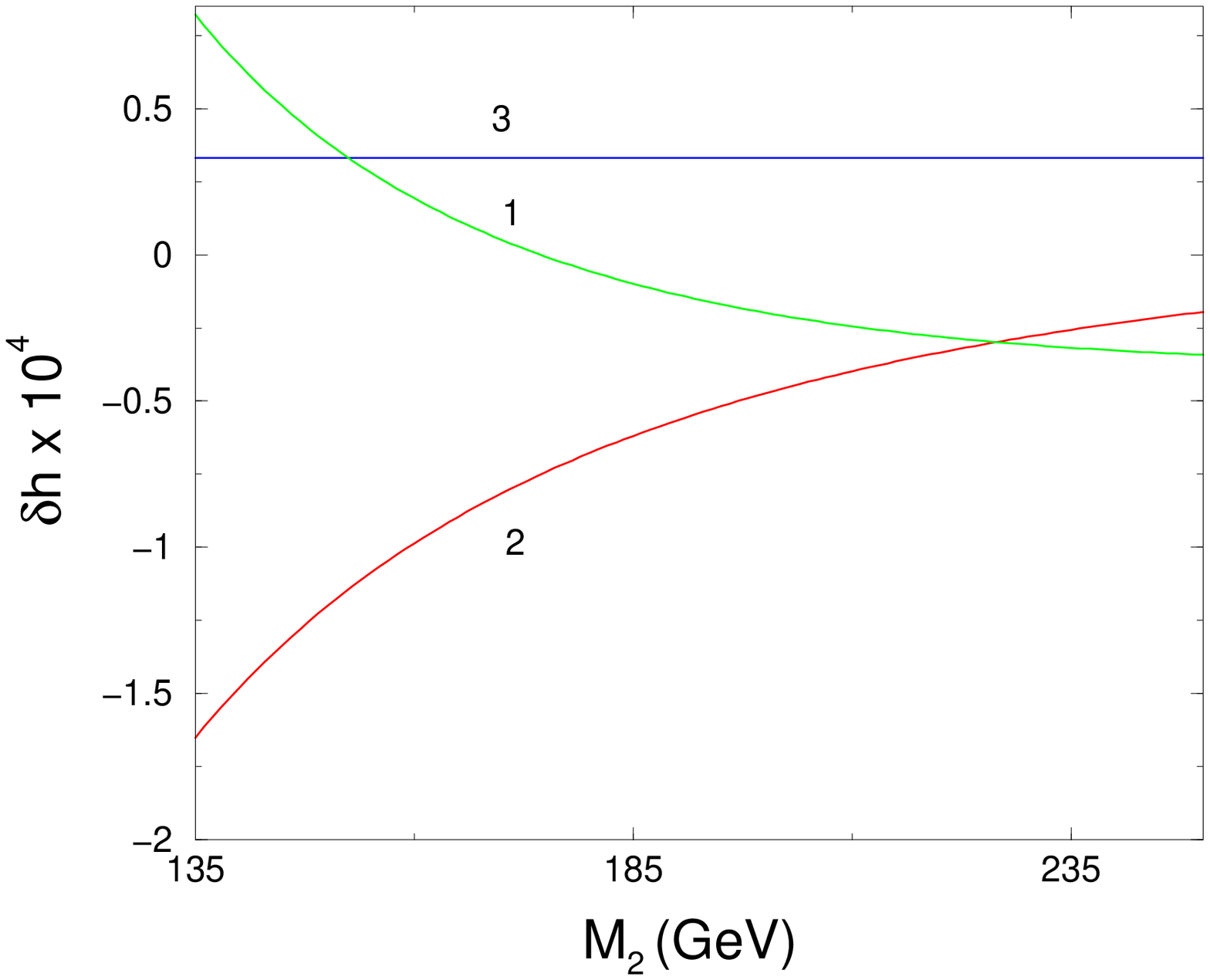, width=11cm, height=8.0cm}\\  
\end{tabular}
\end{center}
\caption{ Vertex corrections to the $Z\bar f f$ couplings as a function
of the GUT scale parameter $M_2$ for $\phi_\mu=0$.
1 -- $\delta h_\nu$, 2 -- $\delta h_{e_L}$, 3 -- $\delta h_{e_R}$.
The other GUT scale parameters are  $m_{\tilde l}=10$ GeV,  $m_{\tilde e}=85$ GeV,
$M_1=100$ GeV, $M_3=200$ GeV, $A=100$ GeV, $\tan\beta=3$. The other
scalar mass parameters are set to $100$ GeV and the CP-phases are set to zero.
 }
\label{coupling}
\end{figure}
\begin{figure}[ht]
\begin{center}
\begin{tabular}{c}
\epsfig{file=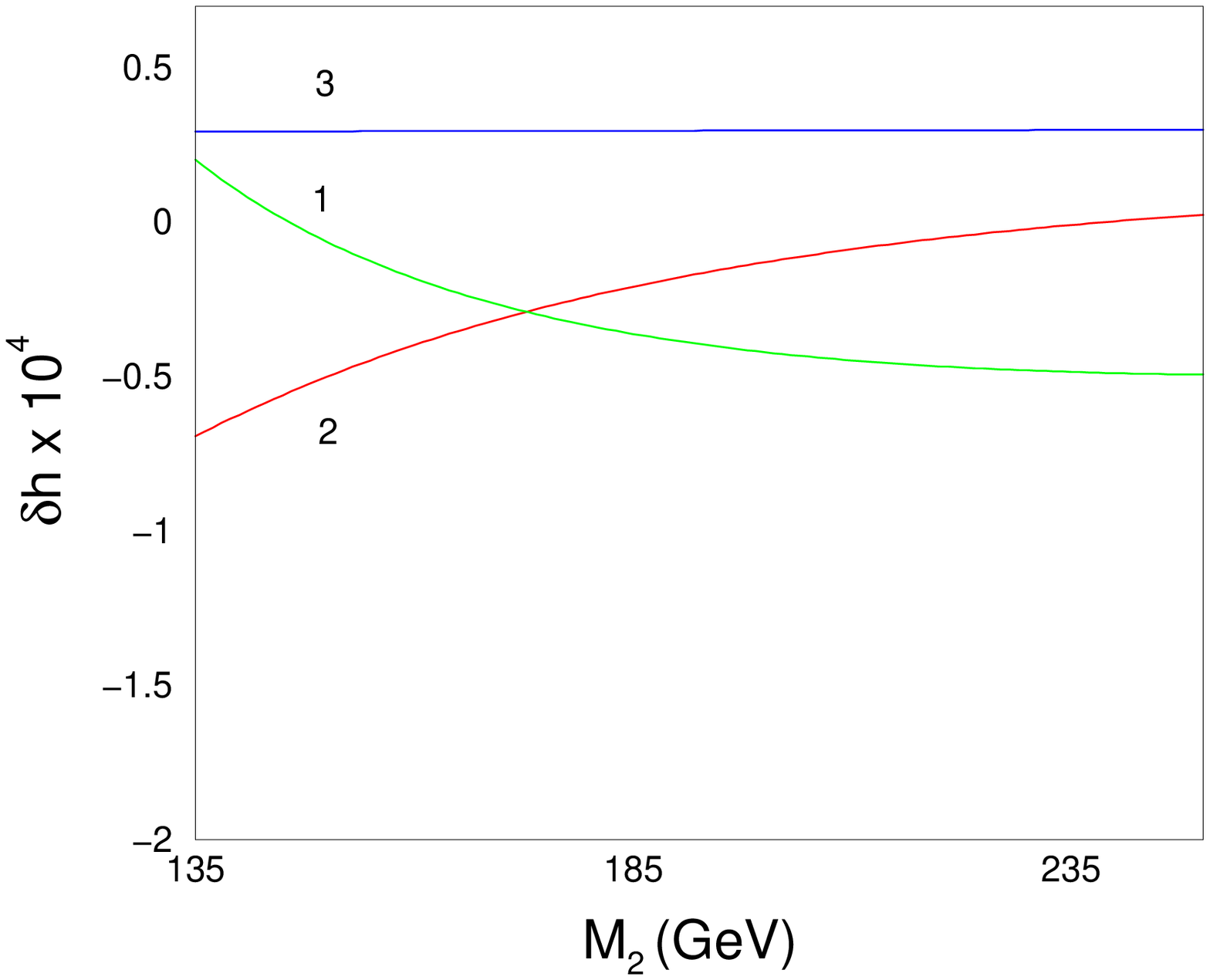, width=11cm, height=8.0cm}\\  
\end{tabular}
\end{center}
\caption{ Vertex corrections to the $Z\bar f f$ couplings as a function
of the GUT scale parameter $M_2$ for $\phi_\mu=\pi$.
1 -- $\delta h_\nu$, 2 -- $\delta h_{e_L}$, 3 -- $\delta h_{e_R}$.
The other GUT scale parameters are  $m_{\tilde l}=10$ GeV,  $m_{\tilde e}=85$ GeV,
$M_1=100$ GeV, $M_3=200$ GeV, $A=100$ GeV, $\tan\beta=3$. The other
scalar mass parameters are set to $100$ GeV and the CP-phases are set to zero.
 }
\label{coupling1}
\end{figure}
\begin{figure}[ht]
\begin{center}
\begin{tabular}{c}
\epsfig{file=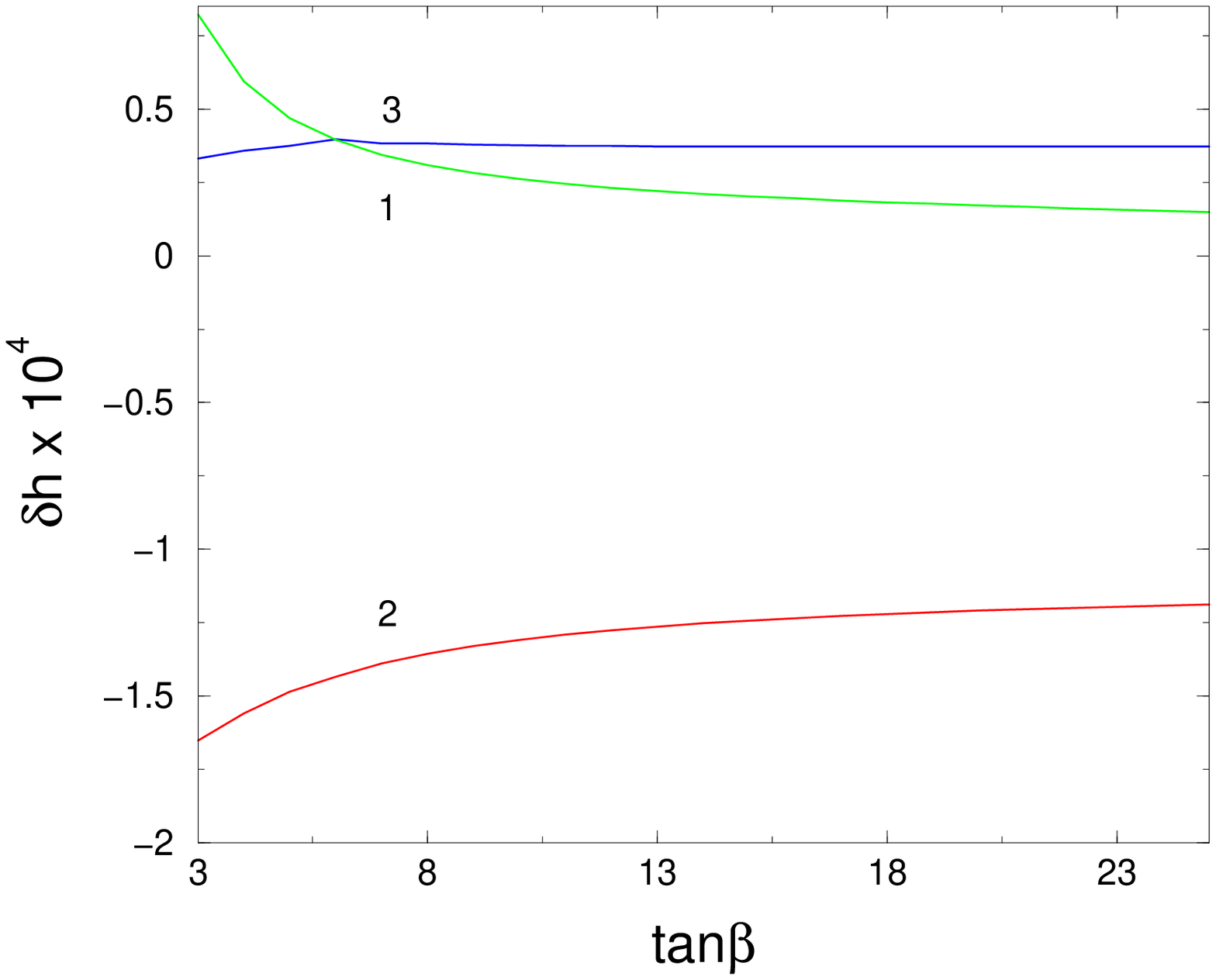, width=11cm, height=8.0cm}\\  
\end{tabular}
\end{center}
\caption{ Vertex corrections to the $Z\bar f f$ couplings as a function
of $\tan\beta$ for $\phi_\mu=0$.
1 -- $\delta h_\nu$, 2 -- $\delta h_{e_L}$, 3 -- $\delta h_{e_R}$.
The other GUT scale parameters are  $m_{\tilde l}=10$ GeV,  $m_{\tilde e}=85$ GeV,
$M_1=100$ GeV, $M_2=135$ GeV,  $M_3=200$ GeV, $A=100$ GeV. The other
scalar mass parameters are set to $100$ GeV and the CP-phases are set to zero.
 }
\label{coupling2}
\end{figure}
\begin{figure}[ht]
\begin{center}
\begin{tabular}{c}
\epsfig{file=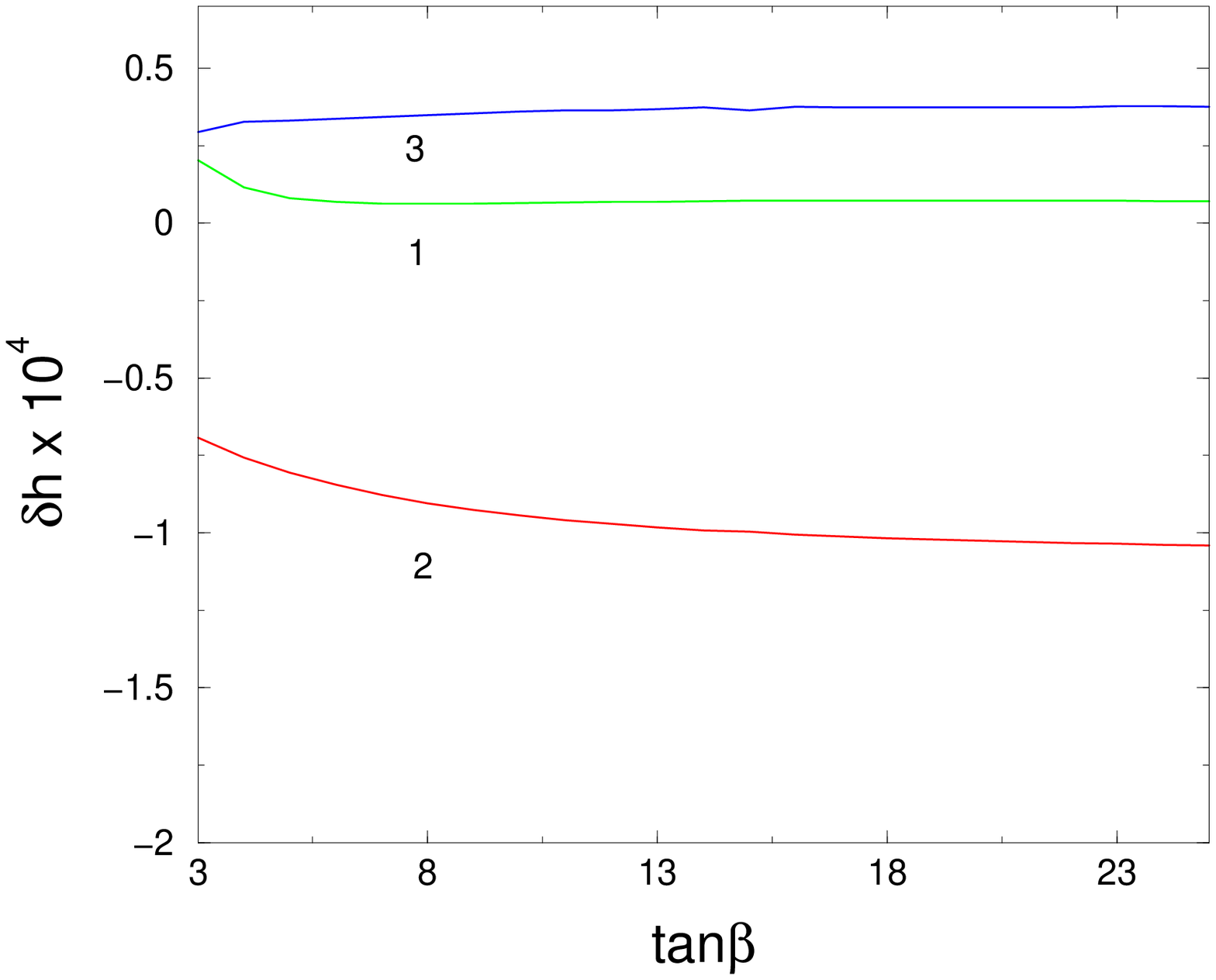, width=11cm, height=8.0cm}\\  
\end{tabular}
\end{center}
\caption{ Vertex corrections to the $Z\bar f f$ couplings as a function
of $\tan\beta$ for $\phi_\mu=\pi$.
1 -- $\delta h_\nu$, 2 -- $\delta h_{e_L}$, 3 -- $\delta h_{e_R}$.
The other GUT scale parameters are  $m_{\tilde l}=10$ GeV,  $m_{\tilde e}=85$ GeV,
$M_1=100$ GeV, $M_2=135$ GeV,  $M_3=200$ GeV, $A=100$ GeV. The other
scalar mass parameters are set to $100$ GeV and the CP-phases are set to zero.
 }
\label{coupling3}
\end{figure}
\begin{figure}[ht]
\begin{center}
\begin{tabular}{c}
\epsfig{file=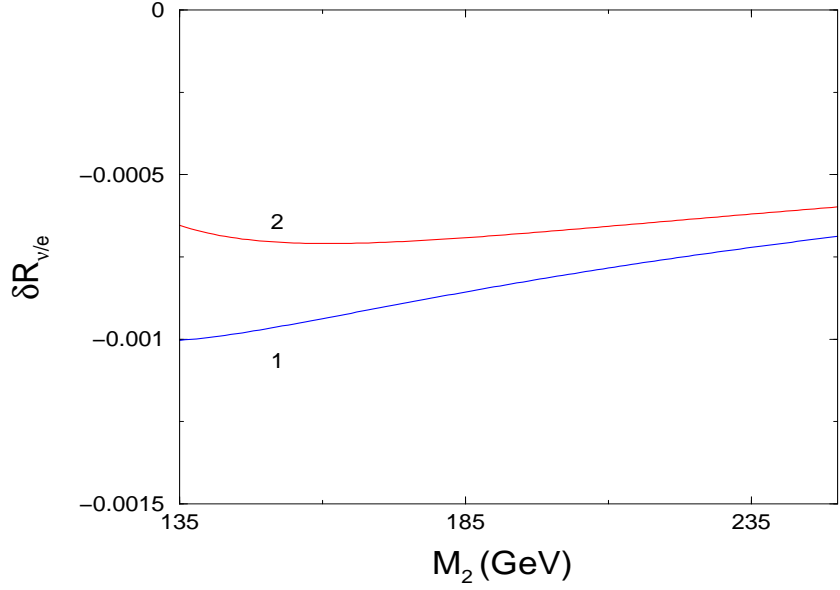, width=11cm, height=8.0cm}\\  
\end{tabular}
\end{center}
\caption{ Shift in  $R_{\nu/e}$ due to the vertex corrections as a function of $M_2$
(this corresponds to the range of the light chargino mass from 95 to 180 GeV).
1 --  $\phi_\mu=0$, 2 --  $\phi_\mu=\pi$.
The  GUT scale parameters are  $m_{\tilde l}=10$ GeV,  $m_{\tilde e}=85$ GeV,
$M_1=100$ GeV, $M_3=200$ GeV, $A=100$ GeV, $\tan\beta=3$. 
 }
\label{R1}
\end{figure}
\begin{figure}[ht]
\begin{center}
\begin{tabular}{c}
\epsfig{file=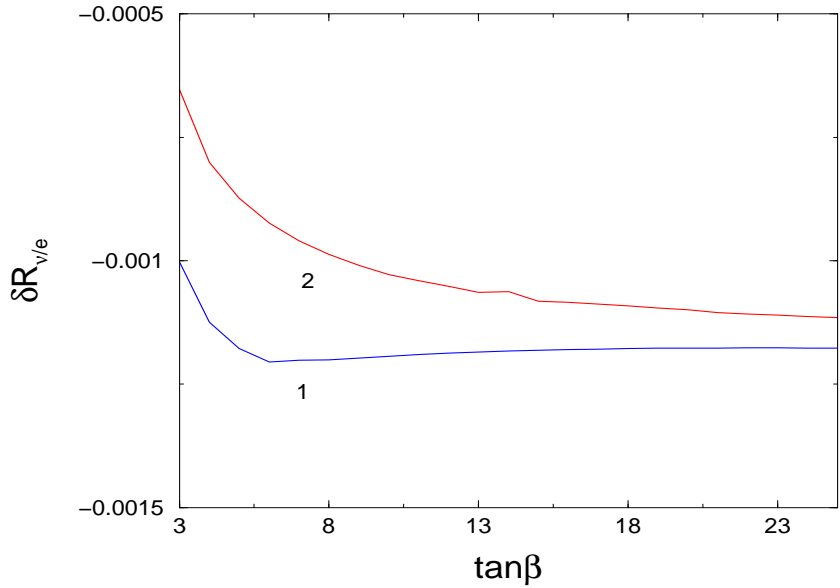, width=11cm, height=8.0cm}\\  
\end{tabular}
\end{center}
\caption{ Shift in  $R_{\nu/e}$ due to the vertex corrections as a function of $\tan\beta$.
1 --  $\phi_\mu=0$, 2 --  $\phi_\mu=\pi$.
The  GUT scale parameters are  $m_{\tilde l}=10$ GeV,  $m_{\tilde e}=85$ GeV,
$M_1=100$ GeV, $M_2=135$ GeV, $M_3=200$ GeV, $A=100$ GeV. 
 }
\label{R2}
\end{figure}
\begin{figure}[ht]
\begin{center}
\begin{tabular}{c}
\epsfig{file=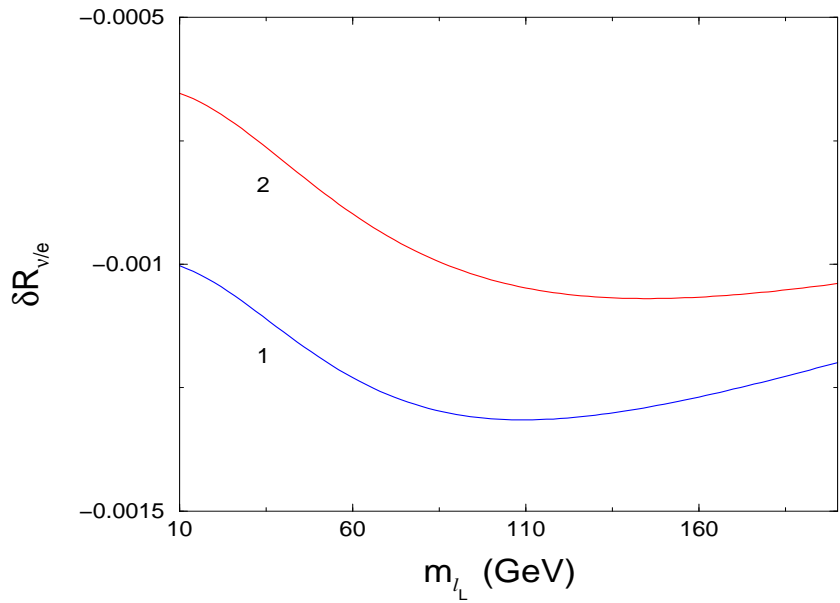, width=11cm, height=8.0cm}\\  
\end{tabular}
\end{center}
\caption{ Shift in  $R_{\nu/e}$ due to the vertex corrections as a function of 
the GUT scale slepton mass parameter $m_{\tilde l}$ (this corresponds to the range of
the slepton mass $m_{\tilde e_L}$ from 104 to 225 GeV).
1 --  $\phi_\mu=0$, 2 --  $\phi_\mu=\pi$. 
The other  GUT scale parameters are    $m_{\tilde e}=85$ GeV,
$M_1=100$ GeV, $M_2=135$ GeV, $M_3=200$ GeV, $A=100$ GeV, $\tan\beta=3$. 
 }
\label{R4}
\end{figure}

\end{document}